\documentclass{aa}  

\usepackage{graphicx}
\usepackage{txfonts}
\usepackage{xcolor}
\usepackage[colorlinks=true,
            linkcolor=blue,
            urlcolor=blue,
            citecolor=blue]{hyperref}
\usepackage{tabularx}

\newcommand{\hmsun}{h^{-1}{\rm M}_\odot}
\newcommand{\hmpc}{h^{-1}{\rm Mpc}}

\usepackage{caption}
\begin{document}   

   \title{The VariableTNG project:}

    \subtitle{Unveiling the physical drivers of galaxy quenching}

   \author{Ignacio G. Alfaro$^{1}$ \thanks{E-mail:ignacio.german@usm.cl}, Antonio D. Montero-Dorta$^{1}$\thanks{E-mail:antonio.montero@usm.cl}, Hong Guo$^{2}$\thanks{E-mail:guohong@shao.ac.cn}, Catalina Riveros-Jara$^{3}$, Kexin Liu$^{2,4}$, Yikai Liu$^{2,4}$, Luis C. Ho$^{5,6}$}
   
   \authorrunning{I. G. Alfaro et al.}
   
   \institute{
              Departamento de Física, Universidad Técnica Federico Santa María, Av. Vicuña Mackenna 3939, San Joaquín, Santiago, Chile. 
              \and
              Shanghai Astronomical Observatory, Chinese Academy of Sciences, Shanghai 200030, China. 
              \and
              Departamento de Informática, Universidad Técnica Federico Santa María, Av. España 1680, Valparaíso, Chile.
              \and
              University of Chinese Academy of Sciences, Beijing 100049, China.
              \and
              Kavli Institute for Astronomy and Astrophysics, Peking University, Beijing 100871, China.
              \and 
              Department of Astronomy, School of Physics, Peking University, Beijing 100871, China.
              }

   \date{\today}

  \abstract
   {Understanding the physical processes that regulate galaxy quenching is a key challenge in galaxy formation and evolution. The VariableTNG (VTNG) project provides an ideal laboratory to investigate these processes, as it systematically varies eight parameters of galaxy formation while keeping the initial conditions fixed, allowing the effects of individual feedback prescriptions to be isolated. We use interpretable machine-learning techniques as a complementary tool to identify the parameters that most strongly regulate the quenched galaxy fraction and to guide the physical interpretation of their effects.}
   {Our goal is to quantify the relative importance of the galaxy formation and feedback parameters that regulate the quenched galaxy fraction at $z=0$, determine how their influence changes across stellar mass and environment.}
   {We compute the quenched galaxy fraction for 26 VTNG realizations as a function of stellar mass, black hole mass, and gas mass, considering both the total galaxy population and separate samples of central and satellite galaxies. We train Random Forest regressors to predict the variation of the quenched fraction relative to the fiducial TNG100-1 model and use SHAP values to quantify both the magnitude and direction of the influence of each galaxy formation parameter.}
    {Our analysis reveals that only a small subset of the VTNG parameters dominates the variance of the quenched fraction. The stellar feedback wind parameter $\kappa_w$ is the primary driver at low stellar masses, while its importance gradually shifts toward AGN-related parameters, particularly $\epsilon_{\rm f, high}$ and $\epsilon_m$, at higher masses. The supernova temperature, $T_{\rm SN}$, also plays an important role at both extremes of the stellar-mass range. This transition persists when the galaxy population is divided into central and satellite systems, although the relative importance of the parameters varies with environment. Inferred trends also indicate how changes in feedback prescriptions affect the quenched fraction. Comparisons with observational measurements further suggest that variations in these feedback parameters may contribute to the discrepancies between the fiducial TNG100-1 model and the observed passive galaxy population.}
   {}
   \maketitle  
%
\section{Introduction}
\label{sec:introduction}

The distribution and evolution of galaxies are shaped by the interplay between hierarchical structure formation and internal baryonic processes \citep[e.g.,][]{Gunn_and_Gott_1972, Ostriker_and_Peebles_1973, White_and_Rees_1978, Mo_Mao_and_White_1998, Springel_and_Hernquist_2003, Di_Matteo_2005, Hopkins_2005, Croton_2006, Somerville_and_Dave_2015}.
Observations from large galaxy surveys have long established a pronounced color bimodality, separating the galaxy population into two distinct groups: actively star-forming galaxies characterized by blue colors and younger stellar populations, and "quenched" galaxies, which are red, passive, and composed of older stellar populations  \citep{Strateva_2001, Blanton_2003, Baldry_2004}. 
The relative abundance of these two populations depends strongly on galaxy mass, environment, and cosmic time, reflecting the variety of physical mechanisms that regulate or suppress star formation \citep[e.g.,][]{Wetzel_2012, Wetzel_2013, Muzzin_2013, Peng_2015, Darvish_2017, Poggianti_2017, Fang_2018}.
Cosmological hydrodynamical simulations have become essential for exploring these processes in a controlled theoretical framework \citep[e.g.,][]{Donnari_2021, Pandey_2025}. 
Models such as IllustrisTNG \citep[TNG; ][]{Naiman_2018, Pillepich_2018a, Springel_2018} have successfully reproduced many observed galaxy properties by incorporating sub-grid prescriptions for radiative cooling, star formation, and feedback \citep{pillepich_2018b}.
Nevertheless, the uncertainties associated with these prescriptions indicate that the full range of physically plausible galaxy growth pathways is still not fully captured \citep[e.g.,][]{pillepich_2018b, Delgado_2023, Villaescusa-Navarro_2021}. 
Moreover, cosmological simulations suffer from significant parameter degeneracies, and different combinations of sub-grid feedback prescriptions can yield similarly realistic galaxy populations at $z=0$, yet imply fundamentally different physical pathways for galaxy evolution \citep[e.g.,][]{pillepich_2018b, Delgado_2023}. 
A critical open problem in galaxy formation is identifying exactly which feedback parameters, and in what combinations, dominate the variance of key observables.
To systematically explore this vast parameter space, the field has increasingly adopted large suites of simulations coupled with machine learning (ML), such as the CAMELS project \citep{Villaescusa-Navarro_2021, Ni_2023} and the FLAMINGO simulations \citep{Kugel_2023, Schaye_2023}. 
Recent studies utilizing these suites have made significant strides; for instance, applying Random Forest (RF) classifiers to the highest-resolution FLAMINGO simulation has definitively highlighted black hole feedback as the primary in situ quenching driver \citep{Lim_2025}. 
Similarly, analyses of the CAMELS suite demonstrate that varying sub-grid parameters within entirely distinct baseline models, for example SIMBA \citep{Dave_2019} versus TNG, produce fundamentally different mappings between feedback efficiencies and global observables like star formation rates and gas fractions \citep{Ni_2023}. 
However, because these broad projects simultaneously vary cosmology and astrophysics across different volumes, or rely on differing hydrodynamical solvers, disentangling the pure, isolated effect of individual feedback parameters remains a complex challenge. 
The VariableTNG \citep[VTNG; described in detail in ][]{liuVTNG} Project was designed to directly address this gap. 
Built on the moving-mesh code \textit{Arepo} \citep{Springel_2010}, VTNG consists of a suite of cosmological magneto-hydrodynamical simulations in which eight parameters governing stellar and active galactic nucleus (AGN) feedback are systematically varied around their fiducial TNG values. 
By keeping the initial conditions exactly fixed across all realizations, VTNG enables a flawless one-to-one comparison between matched galaxies, breaking the degeneracies inherent to broader suites and cleanly isolating the impact of specific feedback variations from cosmic variance.
These parameter variations have a substantial impact on both individual galaxy properties and the global distribution of matter. 
Stellar feedback, controlled by parameters such as the supernova (SN) temperature and wind velocity, regulates disk formation and survival, particularly in low-mass galaxies. 
In contrast, AGN self-regulation and kinetic feedback play a dominant role in shaping the structural evolution and gas depletion of massive systems \citep{Pillepich_2018a}. 
Understanding how these changes affect the galaxy stellar mass function (GSMF) is therefore crucial, since differences between realizations can already appear at $M_{\star} \sim 10^7\ h^{-1}M_{\odot}$ and become increasingly pronounced toward the high-mass end, highlighting the sensitivity of galaxy growth to the adopted feedback model \citep{Ni_2023, Tillman_2023}.
In this context, the quenched galaxy population provides a powerful probe of galaxy formation models \citep[e.g.,][]{Donnari_2021, Pandey_2025}. 
Deep studies of individual fiducial simulation boxes, such as EAGLE \citep{Crain_2015, Schaye_2015}, IllustrisTNG, and SIMBA, have long debated the precise mechanisms driving quenching across different mass scales. 
Recent comparative literature highlights that while SN feedback generally regulates star formation in dwarf galaxies by ejecting gas from shallow potential wells, AGN feedback dictates quenching in massive systems \citep[e.g.,][]{Terrazas_2020, Zinger_2020}. 
Specifically, the central black hole mass has emerged as a highly predictive empirical proxy for quiescence in the high-mass regime across multiple modern simulations \citep{Davies_2020, Ayromlou_2021, Bluck_2024}. 
However, the exact non-linear coupling between SN and AGN feedback at intermediate masses and how different sub-grid parameter choices alter this critical transition regime remains heavily debated. 
Our work uniquely contributes to this problem by utilizing the VTNG framework. By employing interpretable ML techniques, specifically RF regressors and SHapley Additive exPlanations \citep[SHAP;][]{Lundberg_2017}, we can characterize how the relative importance of the physical drivers of quenching shifts from galactic stellar winds to AGN self-regulation across stellar mass and environment. 
This approach allows us not only to capture the magnitude of astrophysical uncertainties \citep{Villaescusa-Navarro_2021, Medlock_2025}, but also to extract the physical direction of the parameter influence, directly bridging the gap between parameterized feedback and the observed passive galaxy population.
In this work, we analyze 26 VTNG simulation boxes at redshift zero to quantify the impact of these sub-grid variations on the stellar mass distribution and on the emergence of passive galaxy populations. 
The paper is organized as follows. In Section \ref{sec:data_and_meth}, we describe the data, the simulation parameters, the galaxy catalogs, and the ML methodology employed in this study. 
Section \ref{sec:quenched_frac} presents the galaxy populations, their specific star formation rates, and the quenched fractions (QF) obtained across the different VTNG boxes. 
In Section \ref{sec:drivers}, we analyze the ML results to identify the main drivers of galaxy quenching, and complement them with physical analysis to assess the role of the detected parameters in galaxy evolution. 
Section \ref{sec:central_and_satellites} examines these results after separating the galaxy population into centrals and satellites. 
Finally, Section \ref{sec:discussion} presents a detailed discussion of our main findings, and Section \ref{sec:conclusions} summarizes our conclusions.
%

\section{Data \& Methods}
\label{sec:data_and_meth}
\subsection{VTNG Simulations}
\label{sec:VTNG_data}
\begin{figure}[h!] 
\centering
\includegraphics[width=0.9\columnwidth]{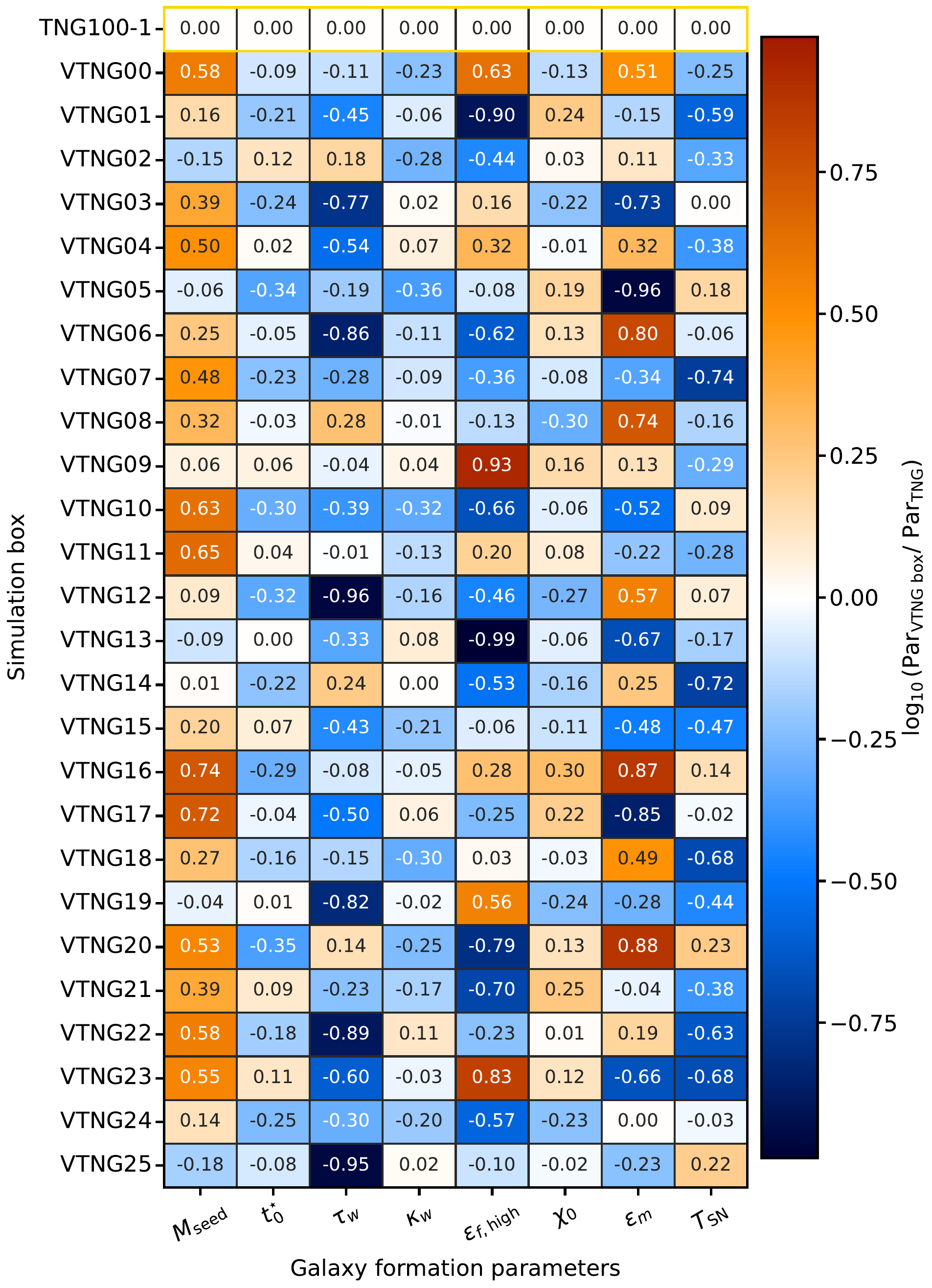}
\caption{Heatmap illustrating the distribution of astrophysical parameters across the 26 VTNG simulation boxes. The color scale of each cell represents the logarithmic ratio $log_{10}(\mathrm{par_{VTNG}/par_{TNG}})$. Rows correspond to individual simulation boxes, and columns correspond to the eight varied parameters.} 
\label{fig:param_heatmap} 
\end{figure}

\begin{table}[t!] 
\centering 
\caption{Galaxy formation and evolution variable parameters in the VTNG simulation suite.} 
\label{tab:vtng_parameters} 
\scriptsize 
\begin{tabular}{lcccc} 
\hline 
Parameter & Symbol & Fiducial (TNG) & Range (VTNG) \\ 
\hline 
Seed black hole mass ($h^{-1} M_{\odot}$) & $M_{\text{seed}}$ & $8\times10^5$ & ($5\times10^5, 5\times10^6$) \\ 
Radio-mode feedback factor & $\epsilon_m$ & 1.0 & (0.1, 10.0) \\ 
Quasar-mode feedback factor & $\epsilon_{f,\text{high}}$ & 0.1 & (0.01, 1.0) \\ 
Quasar threshold & $\chi_0$ & 0.002 & (0.001, 0.004) \\ 
Max star formation timescale (Gyr) & $t^{\star}_0$ & 2.27 & (1.0, 3.0) \\ 
Supernova temperature (K) & $T_{\text{SN}}$ & $5.73\times10^7$ & ($10^7, 10^8$) \\ 
Thermal wind fraction & $\tau_w$ & 0.1 & (0.01, 0.2) \\ 
Wind velocity factor & $\kappa_w$ & 7.4 & (3, 10) \\ 
\hline 
\end{tabular} 
\end{table}
The VTNG project is a dedicated suite of cosmological magnetohydrodynamical simulations designed to isolate the impact of baryonic feedback on galaxy properties under strictly controlled initial conditions.
Built upon the moving-mesh code \textit{Arepo}, VTNG adopts the same overall physical framework as the TNG project. 
The simulations evolve a periodic box of side length $50\,\hmpc$, sampled with $1024^3$ dark matter particles and $1024^3$ gas cells.
This setup yields a dark matter particle mass resolution of $1.28\times10^7\,\hmsun$ and a target baryonic mass resolution of approximately $2.4\times10^6\,\hmsun$. 
The adopted cosmology is consistent with the \cite{Planck_2018} constraints, with parameters $[\Omega_m, \Omega_b, \sigma_8, h] = [0.3158, 0.04939, 0.8120, 0.6732]$.
The sub-grid model includes the main baryonic processes relevant for galaxy formation, namely radiative cooling, star formation, stellar evolution, and feedback from active galactic nuclei (AGN) \citep{ pillepich_2018b}. 
Star formation is modeled through a two-phase interstellar medium prescription, in which cold gas above a density threshold is converted into stars \citep{Springel_and_Hernquist_2003, pillepich_2018b}. 
Stellar feedback is implemented through galactic winds that inject both kinetic and thermal energy into the surrounding medium \citep{pillepich_2018b}.
For supermassive black holes (SMBHs), the model adopts a dual-mode AGN feedback scheme: 
a high-accretion ``quasar'' mode, which deposits thermal energy, and a low-accretion ``radio'' mode that injects momentum and energy episodically into the surrounding gas \citep{Weinberger_2017}.
VTNG systematically varies eight astrophysical parameters that regulate these feedback mechanisms, while keeping the large-scale initial conditions fixed across all realizations. 
Rather than varying each parameter independently, the simulation suite employs a Sobol low-discrepancy sequence to sample the eight-dimensional parameter space, ensuring uniform coverage of the allowed parameter ranges and reducing parameter-wise degeneracies \citep{Sobol_2001}. 
These parameters are: 
\begin{itemize}
    \item Maximum star formation timescale ($t_0^\star$): Sets the timescale for converting gas into stars at the density threshold, thereby regulating the global star formation history.
    
    \item Supernova temperature ($T_{\rm SN}$): Controls the energy injected by SN per unit stellar mass. Larger values suppress star formation more efficiently in low-density environments.
    
    \item Thermal wind fraction ($\tau_w$): Defines the fraction of stellar feedback energy injected as thermal rather than kinetic energy, with a strong impact on the thermal state of the circumgalactic medium \citep{Ni_2023}.
    
    \item Wind velocity factor ($\kappa_w$): Sets the normalization of galactic stellar feedback wind velocities. This is one of the most influential parameters for the abundance of galaxies and the shape of the low-mass end of GSMF \citep{Muratov_2015, Angles_2017}.
    
    \item Seed black hole mass ($M_{\rm seed}$): Specifies the initial mass assigned to newly seeded black holes in halos, primarily affecting black hole growth in lower-mass systems \citep{Ni_2023}.
    
    \item Radio-mode feedback factor ($\epsilon_m$): Scales the energy injected during the low-accretion AGN mode. This parameter is a key driver of quenching in massive galaxies \citep{Ni_2023}.
    
    \item Quasar-mode feedback factor ($\epsilon_{{\rm f, high}}$): Governs the coupling efficiency of thermal energy during the high-accretion phase, influencing the early growth of black holes and the evolution of massive galaxies.
    
    \item Quasar threshold ($\chi_0$): Sets the accretion-rate threshold separating the quasar and radio modes, thereby controlling when strong AGN feedback is activated \citep{Weinberger_2017, Ni_2023}.
\end{itemize}
The fiducial values and ranges of variation for the full VTNG boxes are summarized in Table \ref{tab:vtng_parameters}.
Figure \ref{fig:param_heatmap} presents a heatmap of the distribution of astrophysical parameters across the 26 VTNG realizations used in this work. 
In that figure, the color scale of each cell represents the logarithmic ratio $\log_{10}(\mathrm{par}_{\rm VTNG}/\mathrm{par}_{\rm TNG})$, where $\mathrm{par}_{\rm VTNG}$ is the parameter value in a given realization and $\mathrm{par}_{\rm TNG}$ is the corresponding fiducial TNG value.

\subsection{Galaxy Selection}
\label{sec:galaxy_select}
Halos and subhalos in the 26 VTNG realizations at $z=0$ are identified using the standard Friends-of-Friends (FoF) and \textsc{subfind} algorithms \citep{Springel_2001, Dolag_2009, pillepich_2018b}. 
Since all VTNG runs share identical initial conditions, the large-scale structure evolves coherently across the entire suite, allowing individual objects to be directly compared under different feedback prescriptions. 
This unique feature enables a controlled study of how variations in baryonic physics affect galaxy evolution while minimizing the impact of cosmic variance.
In VTNG, following the standard TNG definitions, the most massive subhalo located at the minimum of the gravitational potential within each FoF halo is identified as the central galaxy, while all remaining gravitationally bound subhalos are classified as satellites \citep{pillepich_2018b}. The galaxy catalog is then constructed in two stages.
First, we perform a cross-identification of galaxies among the 26 VTNG realizations. 
We consider all subhalos with stellar mass $M_{\star}>0$ and match them based on their spatial positions, using a tolerance radius of $1\,\hmpc$. 
Owing to the common initial conditions, this procedure allows us to identify the same underlying objects across different realizations and to construct a matched sample of $50246$ subhalos. 
This sample corresponds to approximately 79\% of the galaxies identified in the realization containing the largest number of objects, without applying other restrictions, such as mass cuts.
Second, we impose a stellar mass threshold to ensure that only well-resolved galaxies are included in the analysis. 
We restrict our study to subhalos with stellar masses $M_{\star}>10^9\,\hmsun$, a limit above which the internal properties and star-formation activity of galaxies are robustly resolved. 
The resulting numbers of total, central, and satellite galaxies in each realization are summarized in Table \ref{tab:galaxy_counts}.

\begin{table}[h!]
\centering
\caption{Galaxy population statistics for the final samples in the TNG100-1 and 26 VTNG realizations at z=0.} 
\scriptsize 
\label{tab:galaxy_counts} 
\begin{tabular}{lccc} 
\hline 
Realization & $N_{gal}$ & $N_{centrals}$ & $N_{satellites}$ \\ 
\hline 
TNG100-1 & 17695 & 9371 (52.96\%) & 8324 (47.04\%) \\ 
VTNG 00 & 4716 & 2986 (63.32\%) & 1730 (36.68\%) \\ 
VTNG 01 & 3857 & 2425 (62.87\%) & 1432 (37.13\%) \\ 
VTNG 02 & 3484 & 2241 (64.32\%) & 1243 (35.68\%) \\ 
VTNG 03 & 5635 & 3165 (56.17\%) & 2470 (43.83\%) \\ 
VTNG 04 & 3377 & 2025 (59.96\%) & 1352 (40.04\%) \\ 
VTNG 05 & 6886 & 3618 (52.54\%) & 3268 (47.46\%) \\ 
VTNG 06 & 4669 & 2896 (62.03\%) & 1773 (37.97\%) \\ 
VTNG 07 & 3556 & 2246 (63.16\%) & 1310 (36.84\%) \\ 
VTNG 08 & 3733 & 2269 (60.78\%) & 1464 (39.22\%) \\ 
VTNG 09 & 3772 & 2238 (59.33\%) & 1534 (40.67\%) \\ 
VTNG 10 & 6008 & 3426 (57.02\%) & 2582 (42.98\%) \\ 
VTNG 11 & 3904 & 2427 (62.17\%) & 1477 (37.83\%) \\ 
VTNG 12 & 6456 & 3551 (55.00\%) & 2905 (45.00\%) \\ 
VTNG 13 & 3810 & 2250 (59.06\%) & 1560 (40.94\%) \\ 
VTNG 14 & 3141 & 1954 (62.21\%) & 1187 (37.79\%) \\ 
VTNG 15 & 3465 & 2229 (64.33\%) & 1236 (35.67\%) \\ 
VTNG 16 & 6464 & 3467 (53.64\%) & 2997 (46.36\%) \\ 
VTNG 17 & 3987 & 2369 (59.42\%) & 1618 (40.58\%) \\ 
VTNG 18 & 3507 & 2295 (65.44\%) & 1212 (34.56\%) \\ 
VTNG 19 & 3784 & 2335 (61.71\%) & 1449 (38.29\%) \\ 
VTNG 20 & 5986 & 3328 (55.60\%) & 2658 (44.40\%) \\ 
VTNG 21 & 3452 & 2182 (63.21\%) & 1270 (36.79\%) \\ 
VTNG 22 & 3075 & 1836 (59.71\%) & 1239 (40.29\%) \\ 
VTNG 23 & 2888 & 1818 (62.95\%) & 1070 (37.05\%) \\ 
VTNG 24 & 5673 & 3373 (59.46\%) & 2300 (40.54\%) \\ 
VTNG 25 & 5781 & 3200 (55.35\%) & 2581 (44.65\%) \\ 
\hline 
\end{tabular} 
\end{table}

\begin{figure*}[h!] 
\centering 
\includegraphics[width=0.77\textwidth]{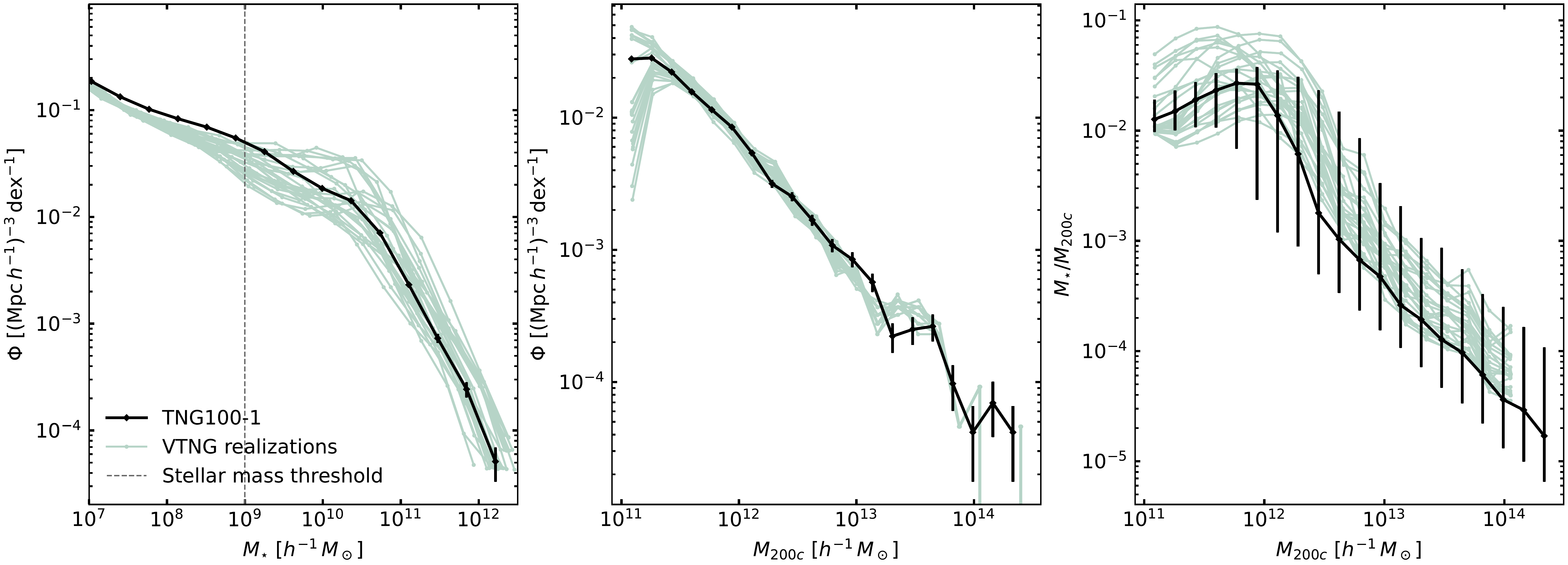}
\caption{Summary of galaxy and halo properties at z=0. \textbf{Left:} Stellar mass functions for the VTNG realizations (gray lines) compared to TNG100-1 (black solid line). The vertical dashed line marks the $M_{\star}=10^9 \hmsun$ threshold. \textbf{Middle:} Halo mass function for the final selected galaxy samples. \textbf{Right:} The ratio of stellar mass to halo mass as a function of halo mass, illustrating the impact of feedback variations on star formation efficiency.} \label{fig:catalogs_summary}
\end{figure*}

The overall properties of the resulting galaxy catalogs are shown in Figure \ref{fig:catalogs_summary}. 
The left panel presents the GSMFs of the 26 VTNG realizations. 
For comparison, we include the fiducial TNG100-1 result. 
The VTNG realizations exhibit significant differences in their GSMFs, illustrating how variations in the feedback parameters modify the efficiency of galaxy formation across a wide range of stellar masses. 
The vertical dashed line indicates the adopted stellar mass threshold of $M_{\star}=10^9\,\hmsun$.
The middle panel shows the halo mass functions (HMFs), demonstrating that the underlying dark matter distribution remains very similar across all realizations, as expected from the shared initial conditions. 
Finally, the right panel presents the stellar-to-halo mass relation, $M_{\star}/M_{\rm halo}$ as a function of $M_{\rm halo}$. 
The dispersion among the different realizations illustrates how variations in the eight astrophysical parameters modulate the efficiency of galaxy formation and star formation within dark matter halos.
%


\subsection{Machine learning framework}
\label{sec:ML_tech}
The exploration of a multi-dimensional parameter space from a suite of cosmological simulations requires methods capable of capturing non-linear relationships and interaction effects between parameters that classical statistical approaches may struggle to detect \citep[e.g.,][]{Piotrowska_2022, Jo_2023, Tillman_2023, Lee_2024}.
In this work, we employ a combination of supervised ML and model-agnostic interpretability tools to quantify how variations in the galaxy formation parameters affect the quenched galaxy fraction.
The ML framework is used here as an exploratory and interpretive approach, rather than as an end in itself or as a predictive model. 
Its purpose is to identify systematic trends and parameter dependencies that can then be investigated and interpreted from a physical perspective.
We organize this analysis in three parts. 
In this section, we introduce the main concepts underlying the ML and SHAP framework and describe how they are used to identify parameter dependencies. 
In Section \ref{sec:feed_param}, we apply this methodology to the simulation suite and present the resulting trends.
Finally, the design and performance of the adopted ML models, together with additional validation tests, are presented in Appendix~\ref{app}.
In what follows, we describe the methodological framework of our ML/SHAP analysis.
\begin{itemize}
    \item Random Forest Regressor \citep{Breiman_2001}. The core predictive model used in our analysis. An RF is an ensemble method that constructs a large number of decorrelated decision trees, each trained on a bootstrap resample of the training set and using a random subset of features at every node split. The final prediction is obtained by averaging the outputs of the individual trees, allowing the model to capture non-linear relationships and parameter interactions while reducing the variance of individual trees. This makes RF well suited to our setting, the relationships between parameter variations and observable outputs are expected to be non-linear, and the method requires minimal hyperparameter tuning. We tested different numbers of decision trees and adopted a forest of 300 trees, which provides stable and converged predictions (see App. \ref{app}).
    \item Leave-One-Out cross-validation. ML models are commonly developed using separate training, validation, and test sets. However, because the number of simulations is comparable to the number of free parameters, a split would leave either the training or the evaluation sets severely underpopulated, reducing both the amount of information available for learning and the robustness of the performance estimates.
     To maximize the use of our simulation suite, we therefore adopt Leave-One-Out cross-validation \citep[LOO-CV;][]{Stone_1974}, in which the model is retrained $N-1$ times, each time holding out a single simulation and predicting its output. The resulting out-of-sample predictions are collected and used to compute the LOO coefficient of determination:
    \begin{equation}
    \label{eq:R}
        R^2_{\rm{LOO}} = 1 - \frac{\sum_i(y_i - \hat{y_i})^2}{\sum_i(y_i - \bar{y_i})^2} 
    \end{equation}
    where $y_i$ are the normalized targets and $\hat{y}_i$ the LOO predictions. This provides an estimate of the model's generalization performance without data leakage, and is used as a diagnostic to identify where the RF has limited predictive power. The dependence of the model performance on the number of simulations included in the LOO-CV procedure is examined in App. \ref{app}.
    \item SHAP Feature Importance. To interpret the predictions of each RF model, we employ SHAP (SHapley Additive exPlanations), a model-agnostic framework based on Shapley values from cooperative game theory. For tree-based models, SHAP values can be computed efficiently using the TreeExplainer algorithm \citep{Lundberg_2019}. For each RF model, SHAP assigns a signed contribution to every input parameter for each simulation, quantifying how that parameter contributes to the corresponding prediction relative to the model baseline. We use the mean absolute SHAP value across simulations to quantify the global importance of each parameter within each RF model.
    \item Sign of the SHAP Influence. The mean absolute SHAP value quantifies the magnitude of the parameter contribution but removes its sign. To characterize the overall direction of this influence, we compute the Spearman rank-correlation coefficient $\rho_s$ between the parameter deviations and their corresponding signed SHAP values across the simulation suite. The Spearman correlation is preferred over the Pearson coefficient because it is invariant to monotonic non-linear transformations and robust to outliers \citep{Press_2007, Feigelson_2012}. The signed importance is then defined as:
    \begin{equation}
        \label{eq:sign}
        \phi_{j,k}^{\pm} = |\phi_{j,k}| \operatorname{sign}(\rho_s)
    \end{equation}
    where, for each mass bin, $|\phi_{j,k}|$ is the mean absolute SHAP value for parameter $j$ in the element $k$ and $\rho_s$ is evaluated from the joint distribution of $(Y_j, \phi_{j})$ over the simulation suite. This construction allows the importance to convey both the magnitude and the direction of each parameter's effect.        
\end{itemize}

\section{Quenched fractions}
\label{sec:quenched_frac}
\begin{figure}[h!] 
\centering
\includegraphics[width=0.95\columnwidth]{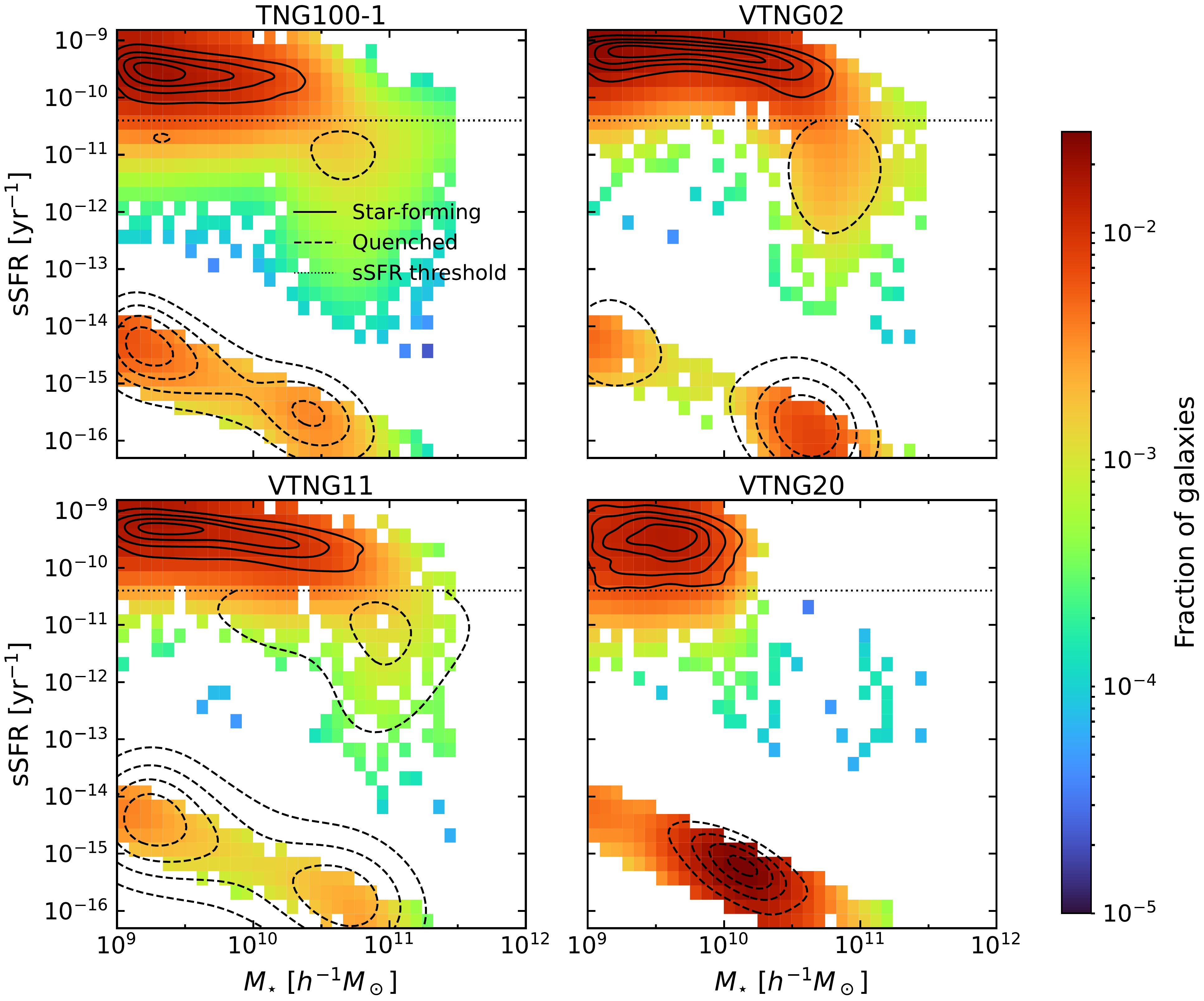}
\caption{Distributions of star-forming (black solid lines) and quenched (black dashed lines) galaxies in the sSFR--stellar mass plane. The top left panel corresponds to the reference TNG100-1 galaxy catalogs, the rest to different VTNG realizations randomly selected. The horizontal dotted line shows the sSFR threshold to separate both samples.} 
\label{fig:ssfr_threhols} 
\end{figure}

Once the galaxy samples were constructed for each VTNG realization (Sec. \ref{sec:galaxy_select}), we investigated the properties of quenched galaxies and their dependence on the different feedback prescriptions. 
Following standard definitions adopted in the literature, galaxies are classified as quenched according to their specific star formation rates (sSFR) \citep[e.g.,][]{Fontanot_2009,Wetzel_2013, Sherman_2021, Donnari_2021, Ni_2023}. 
Throughout this work, a galaxy is considered quenched if $\log_{10}(\text{sSFR/yr}^{-1}) < -10.4$, based on a visual criterion applied to the TNG100-1 reference sample.
To account for the finite numerical resolution of the simulations, galaxies with intrinsic ${\rm SFR}=0$ are assigned a random star formation rate uniformly distributed between $10^{-6}$ and $10^{-5}\,M_{\odot}\,{\rm yr}^{-1}$ \citep{Springel_and_Hernquist_2003, Donnari_2021}. 
This procedure prevents numerical divergences in the logarithmic sSFR and ensures that completely passive systems remain properly represented within the quenched population.
The distribution of galaxies in the sSFR--stellar mass plane is shown in Figure \ref{fig:ssfr_threhols}. 
Each panel corresponds to a different VTNG realization, while the upper-left panel displays the fiducial TNG100-1 simulation. 
The adopted quenching threshold is indicated by the horizontal dotted line, separating the star-forming and passive populations. 
The figure illustrates that the relative abundance of both populations varies considerably among the different realizations.
\begin{figure*}[h!] 
\centering
\includegraphics[width=0.78\textwidth]{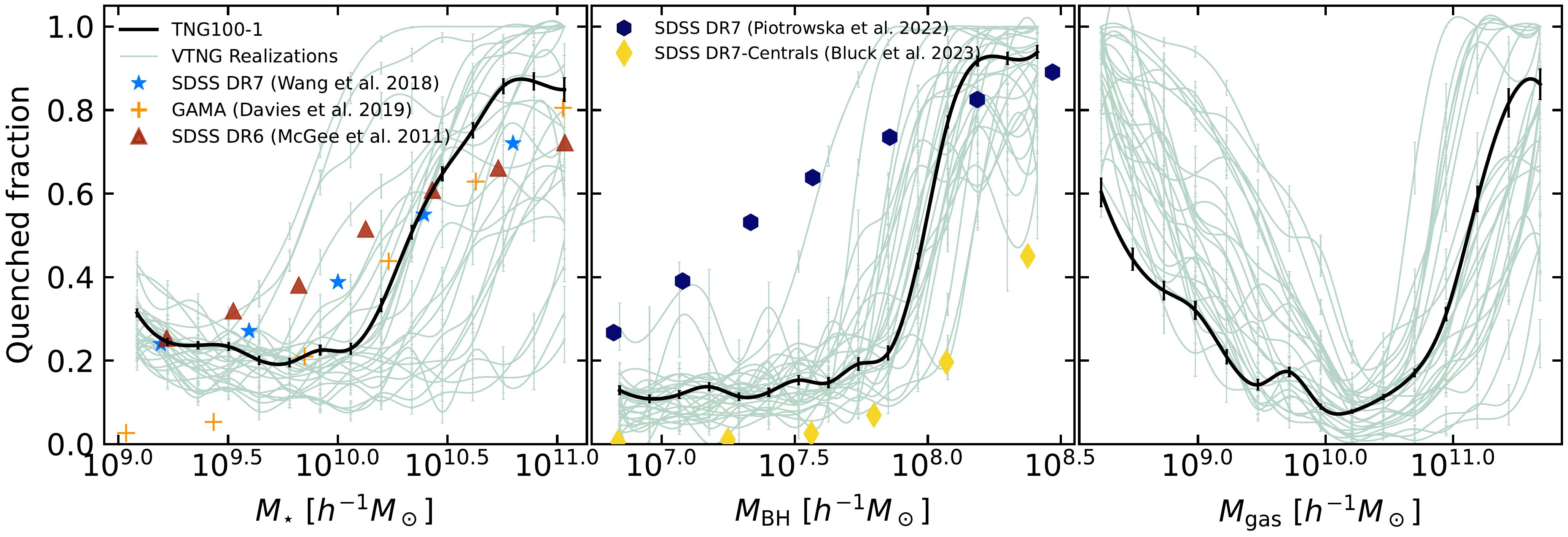}
\caption{Quenched galaxy fraction as a function of the different total masses of a galaxy. All the lines have been smoothed, and the error bars indicate the standard deviation across the sample. The gray lines correspond to VTNG realizations and the black solid line to the reference TNG100-1 simulation. \textbf{Left:} Total stellar mass. The observational measurements are shown as blue stars (SDSS DR7), orange crosses (GAMA), and red triangles (SDSS DR6 field). \textbf{Middle:} Total black hole mass. The observational measures are shown as dark blue octagons (SDSS DR7) and yellow diamonds (SDSS DR7 centrals). \textbf{Right:} Total gas mass.} 
\label{fig:qgf_results} 
\end{figure*}
To quantify these differences, Figure \ref{fig:qgf_results} shows the QF as a function of stellar mass (left), black hole mass (middle), and total gas mass (right).
The black curve shows the fiducial TNG100-1 simulation, while the gray curves correspond to the 26 VTNG realizations. 
The relations are smoothed for clarity, and the error bars indicate the standard deviation of the reference sample.
The left panel also compares the simulations with observations from SDSS DR7 \citep[blue stars,][]{Wang_2018}, GAMA \citep[orange crosses,][]{Davies_2019}, and the SDSS DR6 field sample \citep[red triangles,][]{McGee_2011}. 
These surveys probe nearby galaxies ($z \sim 0.06$--$0.2$) and provide suitable observational benchmarks, although they adopt different criteria to identify quenched galaxies.
\cite{Wang_2018} and \cite{McGee_2011} use fixed sSFR thresholds of $\log_{10}({\rm sSFR}/{\rm yr}^{-1}) < -10.5$ and $-11$, respectively, whereas \cite{Davies_2019} adopts a stellar-mass-dependent SFR threshold following \cite{Woo_2013}. 
Overall, the observational estimates show varying levels of agreement with TNG100-1 across the stellar mass range, as explored in detail in Appendix \ref{app:observation}.
The VTNG realizations show substantial variation in the stellar mass dependence of the quenched fraction, indicating a strong sensitivity to the adopted feedback prescriptions. 
In contrast, the relation with black hole mass is considerably more stable, with the quenched fraction increasing monotonically with $M_{\rm BH}$ across nearly all realizations.
Although the scatter among the VTNG models generally increases toward higher masses, it is noticeably smaller when the relation is expressed in terms of black hole mass. 
This suggests that black hole growth provides a more direct tracer of quenching than stellar mass alone \citep{Bluck_2020, Terrazas_2020, Piotrowska_2022}.
The middle panel also includes two observational estimates from SDSS DR7 \citep{Piotrowska_2022,Bluck_2023}, both based on nearby galaxies ($0.02 < z < 0.2$) but adopting different definitions of quenching.
\cite{Piotrowska_2022} use a fixed threshold of $\log({\rm sSFR}/{\rm yr}^{-1}) < -11$, whereas \cite{Bluck_2023} identify central quenched galaxies as those forming stars at a rate more than one order of magnitude below the star-forming sequence at the redshift of each galaxy. 
For \cite{Piotrowska_2022}, we show the average result obtained from their five different black hole mass calibrations, while for \cite{Bluck_2023} we use their estimate based on the gravitational potential, $\phi$, converted into $M_{\rm BH}$ through the $\phi$--$M_{\rm BH}$ relation calibrated from the TNG model (see original work for more context). 
The two observational estimates show noticeable differences from each other and from the simulations. 
These discrepancies, together with the substantial uncertainties in observational estimates of $M_{\rm BH}$, make quantitative comparisons with simulations challenging.
Finally, the right panel shows the quenched fraction as a function of total gas mass. 
Since a homogeneous observational estimate of the total gas reservoir is difficult to obtain, we do not include observational constraints in this panel. 
Given these limitations, we consider all three mass dependencies throughout this work, while focusing more extensively on the stellar mass dependence, for which more robust observational benchmarks are available.
Overall, the large dispersion observed among the VTNG realizations demonstrates that galaxy quenching is highly sensitive to the underlying sub-grid physics.
In the following section, we use ML techniques to identify which of these parameters are the primary drivers of the observed variations in the quenched fraction.
%

\section{Drivers of galaxy quenching}
\label{sec:drivers}
%

\subsection{Random Forest and SHAP Analysis}
\label{sec:feed_param}
\begin{figure*}[h!] 
\centering
\includegraphics[width=0.75\textwidth]{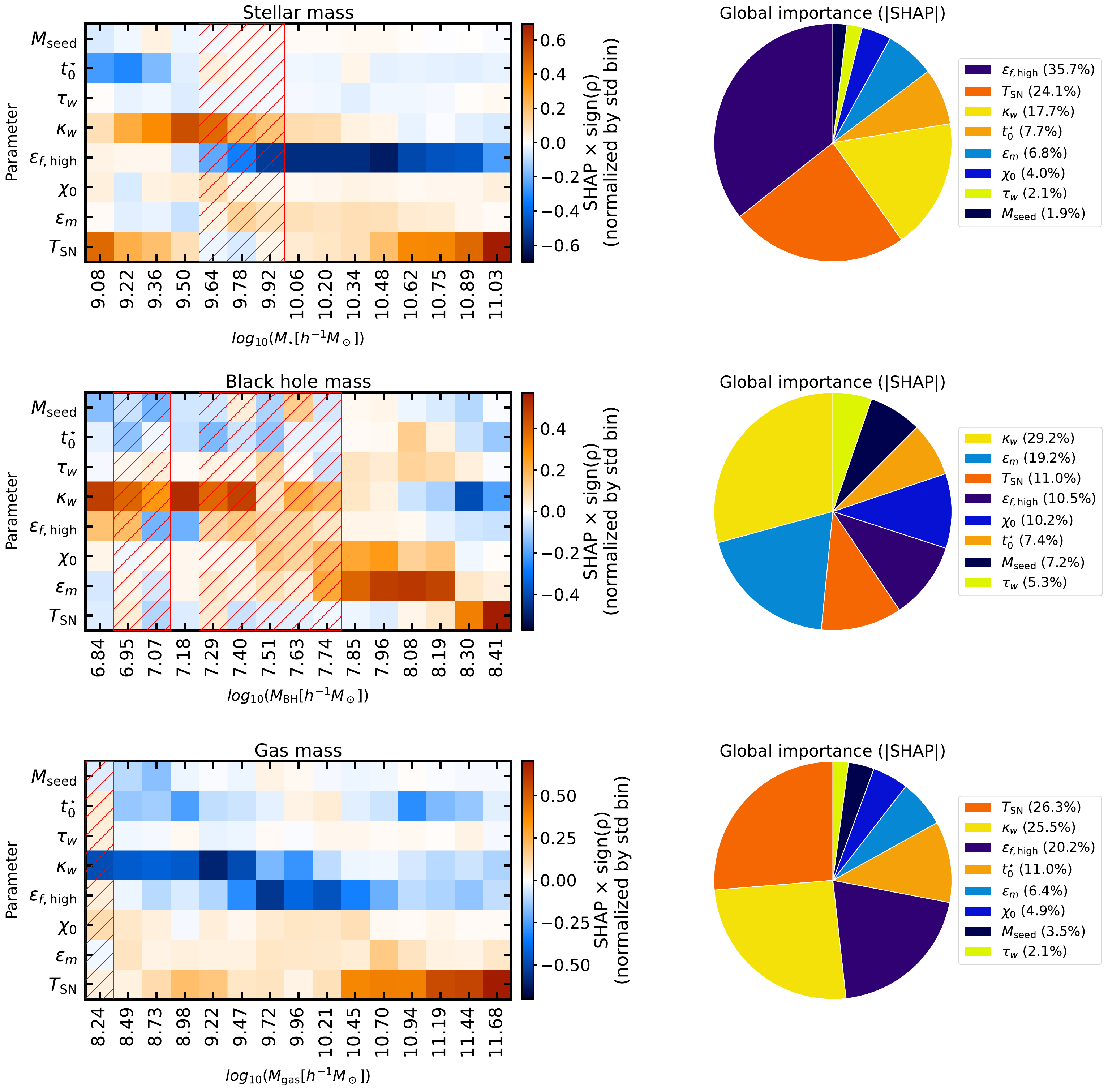}
\caption{SHAP analysis derived from the RF models of the quenched galaxy fraction across VTNG realizations. \textbf{Left:} Heatmaps showing the SHAP value of the 8 astrophysical parameters as a function of different total masses of galaxies. The blue color represents a negative influence of the parameter on the quenched galaxy fraction, and the orange a positive one. Bins marked with red have $R_{\rm LOO}^2 < 0.2$. \textbf{Right:} Pie charts illustrating the global importance of each parameter; values are obtained as the unweighted mean of the absolute SHAP value across the mass bins and the subsequent normalization. \textbf{Top: }quenched fraction as a function of total stellar mass. \textbf{Middle:} quenched fraction as a function of total black hole mass. \textbf{Bottom: }quenched fraction as a function of total gas mass of galaxies.} 
\label{fig:rf_importance} 
\end{figure*}
To quantify the relationship between the astrophysical parameters varied in the VTNG suite and the resulting statistical properties of the galaxy population, we constructed a simulation-level dataset based on differences relative to the fiducial TNG100-1 reference simulation. 
We defined the output variable of our model as the difference in the quenched galaxy fraction ($\Delta \rm{QF}$) within each stellar mass bin for every VTNG realization compared to the reference. 
This vector-based approach captures how variations in the sub-grid physics modify the efficiency of star formation cessation across different mass scales.
In parallel, for each simulation, we defined an input vector composed of the eight physical parameters expressed as logarithmic differences relative to their fiducial values, consistent with the scaling shown in Fig. \ref{fig:param_heatmap}. 
This formulation allows us to isolate the structural and evolutionary response of the galaxy population directly to the feedback perturbations.
The analytical problem was formulated as a series of independent regression models, with one RF trained for each stellar mass bin.
We also tested Ridge regression and Gaussian Process Regression (GPR), finding consistent global importance patterns across the three approaches; we therefore adopted the RF as our fiducial model because of its ability to capture nonlinear relationships while providing a robust estimate of parameter importance.
Before fitting, the target in each mass bin was standardized by its standard deviation, ensuring that all bins contribute equally to the subsequent importance analysis regardless of the absolute amplitude of their variation. 
The predictive performance of each RF model was evaluated using the LOO-CV technique (Sec. \ref{sec:ML_tech}), yielding a coefficient of determination $R_{\rm{LOO}}^2$ (Eq. \ref{eq:R}) for each mass bin. 
This provides a bin-dependent measure of the predictive reliability of the models.
Afterwards, each RF was interpreted independently using SHAP. 
Thus, each mass bin has its own set of SHAP values, providing the signed contribution of each parameter (Eq. \ref{eq:sign}) to the predicted $\Delta \mathrm{QF}$ for that bin.
The dependence of the results on the number of decision trees is analyzed in App. \ref{app}, showing that this choice does not have a significant impact.
Further details of the RF/SHAP methodology and the construction of the importance metrics are provided in Appendix \ref{app:rf_method}.
The resulting signed parameter importances are presented in Figure \ref{fig:rf_importance} as heatmap. 
Each column corresponds to an independent RF/SHAP analysis for a given stellar mass bin, with positive (negative) values indicating that increasing the corresponding parameter tends to increase (decrease) $\Delta \mathrm{QF}$. 
Bins marked with red have $R_{\rm LOO}^2 < 0.2$, indicating lower predictive reliability and therefore requiring caution in interpreting their SHAP-based parameter contributions.
The complete process was repeated using the $\Delta \mathrm{QF}$ as a function of the black hole (medium heatmap) mass and the gas mass (lower heatmap), to explore how the driver parameters change when considering a different component. 
To obtain a single scalar summary of each parameter's overall influence across the full mass ranges, the per-bin absolute SHAP values are aggregated via an unweighted mean over bins.
These global importance values provide a single ranking of the overall influence of each galaxy formation parameter and are shown as the pie charts accompanying the heatmaps in Fig. \ref{fig:rf_importance}. 
The precise definition of the global importance is given in App. \ref{app:rf_method}.
These analyses reveal a clear mass dependence in the influence of the galaxy formation parameters on the QF.
Fig. \ref{fig:rf_importance} shows that this behavior is remarkably consistent across the three mass proxies considered. 
At low masses, galaxy quenching is driven primarily by stellar feedback, with the stellar feedback wind velocity factor $\kappa_w$ emerging as the dominant parameter. 
Toward higher masses, the influence of $\kappa_w$ declines and quenching becomes increasingly regulated by AGN feedback parameters, particularly $\epsilon_{\rm f, high}$ and $\epsilon_m$, together with the SN temperature, $T_\mathrm{SN}$. 
This transition is consistent with the expected shift from stellar and SN feedback to AGN-dominated quenching as galaxies and their central black holes evolve \citep{Silk_2012, Bower_2017, Weinberger_2017, Piotrowska_2022}.
The global importance rankings reinforce this picture. 
Although the relative ordering depends on the chosen mass proxy, the variance in $\Delta \mathrm{QF}$ is consistently dominated by the same small set of parameters. 
In contrast, $\tau_w$, $M_\mathrm{seed}$, and $\chi_0$ contribute only marginally in most cases. 
Overall, the recurrence of the same dominant parameters across the three mass definitions suggests that stellar mass, black hole mass, and gas mass trace different stages of a common feedback-regulated evolutionary sequence, with their differing rankings reflecting the relative importance of stellar and AGN feedback along that evolution.
\subsection{Physical influence of the main parameters}
\label{sec:robustness}
\begin{figure*}[h!] 
\centering
\includegraphics[width=0.85\textwidth]{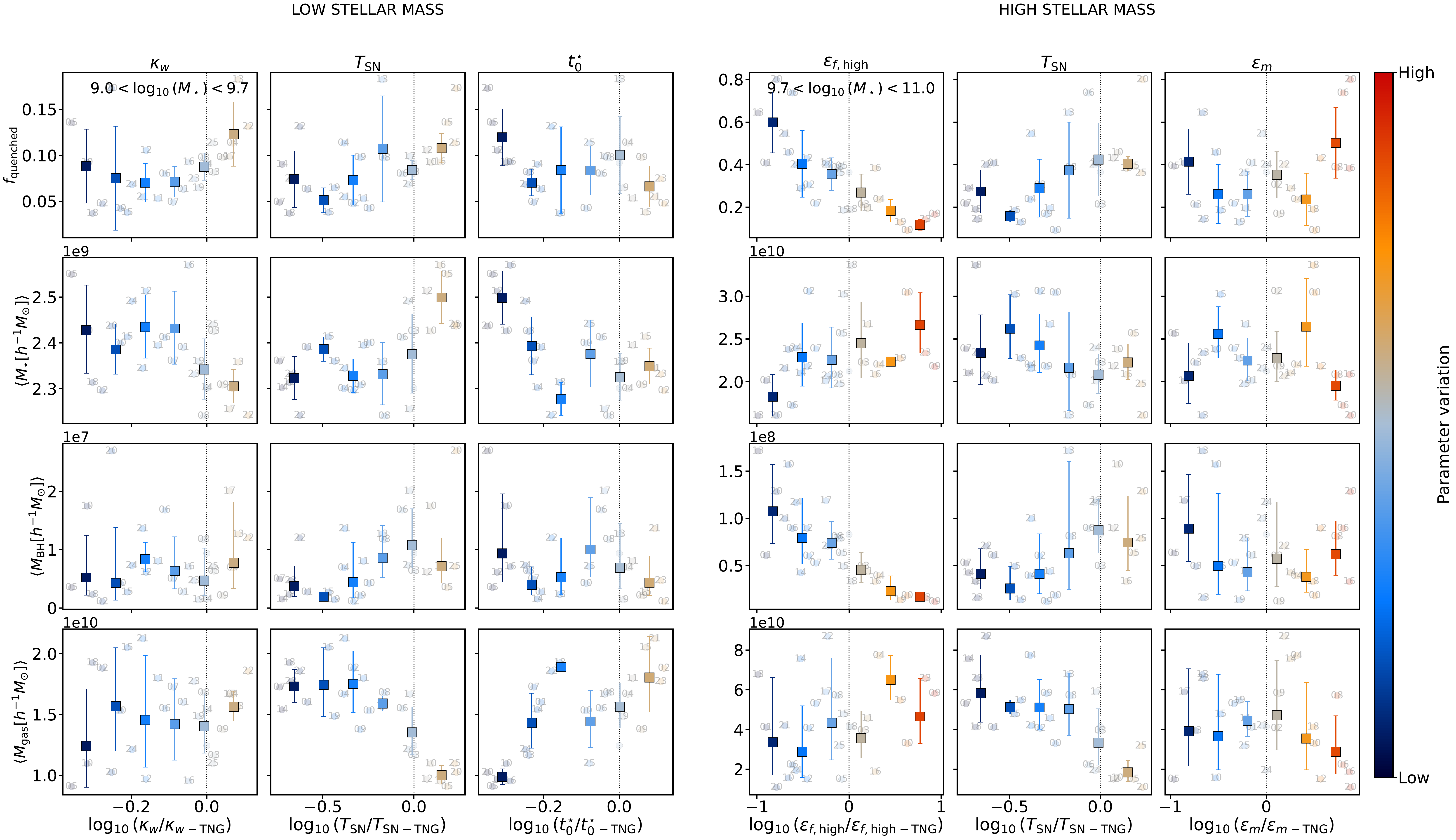}
\caption{Scatter plot of the relationship between the logarithmic variation of main parameters in each VTNG simulation and different properties in their galaxies. Each column of panels corresponds to a single parameter, while each row corresponds to a specific galaxy property. The light semi-transparent circle shows the mean value of each property in a VTNG realization; the number indicates the matching realization. The filled squares with error bars indicate the mean and dispersion within six bins of the logarithmic deviation. \textbf{Left}: columns to the three main parameters detected in the SHAP analysis for the low stellar mass galaxies, $\log_{10}(M_\star\,[h^{-1}M_\odot]) < 9.7$, these are ordered from left to right following their importance and are $(\kappa_w; \ T_{\rm{SN}}; \ t_{0}^{\star})$. \textbf{Right}: columns to the three main parameters detected in the SHAP analysis for the high stellar mass galaxies, $9.7 < \log_{10}(M_\star\,[h^{-1}M_\odot]) < 11.0$, these are ordered from left to right following their importance and are $(\epsilon_{\rm f,high}; \ T_{\rm{SN}}; \ \epsilon_{m})$. \textbf{Top:} Mean quenched galaxy fraction. \textbf{Middle-Top:} Mean stellar mass of galaxies. \textbf{Middle-Bottom:} Mean black hole mass of galaxies. \textbf{Bottom:} Mean gas mass of galaxies.}
\label{fig:scatter} 
\end{figure*}

To complement the ML analysis presented in the previous subsection, we now examine the direct response of galaxy properties to the dominant parameters identified by the RF/SHAP analyzes. 
Unlike the previous analysis, which considered stellar mass, black hole mass, and gas mass as alternative galaxy classifications, here we focus exclusively on stellar mass selected galaxy populations in order to study the underlying physical trends. 
Consequently, only the results presented below can be directly compared with the stellar mass results shown in the top panels of Fig. \ref{fig:rf_importance} to provide a physical interpretation of the inferred parameter dependence.
We split the galaxy population of each simulation in the VTNG suite into two stellar mass regimes: low-mass galaxies, defined as $\log_{10}(M_\star\,[h^{-1}M_\odot]) < 9.7$, and high-mass galaxies, defined as $9.7 < \log_{10}(M_\star\,[h^{-1}M_\odot]) < 11.0$. 
This division follows the transition identified by the RF/SHAP analysis, where the dominant drivers of quenching shift from stellar and SN feedback at low masses to AGN feedback at high masses.
For each simulation, we recompute the quenched fraction as defined in Sec. \ref{sec:quenched_frac} and study its dependence on the parameters identified by the ML analysis as the dominant regulators in each regime:
$\kappa_w$, $T_\mathrm{SN}$, and $t_0^{\star}$ for low stellar mass galaxies, and $\epsilon_{\rm f,high}$, $T_\mathrm{SN}$, and $\epsilon_m$ for high stellar mass galaxies. 
This direct comparison also allows us to verify that the statistical importance assigned by the ML model corresponds to coherent, monotonic physical trends within the simulation suite, rather than to artifacts of the regression procedure \citep{Medlock_2025}.
To interpret these trends, we also examine how the mean stellar mass, black hole mass, and mean gas mass respond to variations in the same parameters, since these quantities provide the physical channels through which stellar and AGN feedback regulate star formation.
Since black hole growth and gas content are the primary channels through which $\kappa_w$, $T_\mathrm{SN}$, $\epsilon_{\rm f,high}$, and $\epsilon_m$ are expected to regulate star formation, this comparison provides a physical link between the parameter-driven response identified by SHAP and the quenching mechanism itself.
Figure~\ref{fig:scatter} summarizes the results. 
The left and right blocks correspond to the low- and high- stellar mass regimes, respectively, while each column shows the dependence on one galaxy formation parameter.
The four rows display, from top to bottom, the quenched fraction, the mean stellar mass, the mean black hole mass, and the mean gas mass. 
Individual simulations are shown as semi-transparent circles, whereas filled squares with error bars indicate the mean and dispersion measured in bins of parameter variation.
In both stellar mass regimes, the top row of Fig. \ref{fig:scatter} confirms the main trends identified in Sec. \ref{sec:feed_param}, in agreement with the stellar mass heatmaps shown in Fig. \ref{fig:rf_importance} (only the top panel). 
For low stellar mass galaxies, increasing $\kappa_w$ and $T_\mathrm{SN}$ both increase the quenched fraction, according to the RF/SHAP analysis, whereas larger values of $t_0^{\star}$ have the opposite effect. 
At high stellar masses, the strongest dependence is found for $\epsilon_{\rm f,high}$, whose increase reduces both the quenched fraction and its scatter across the VTNG realizations. 
These results provide an independent validation that the dominant parameters identified by the ML analysis correspond to genuine physical trends.
The remaining rows of Fig. \ref{fig:scatter} are not part of the RF/SHAP analysis but help to interpret the physical origin of the inferred trends by showing the response of the mean stellar, black hole, and gas masses. 
For low stellar mass galaxies, none of the dominant parameters shows a significant correlation with black hole mass, reinforcing the picture that AGN activity plays only a minor role in this regime.
This behavior is partly explained by the relatively low black hole masses in this regime, where \(M_{\rm BH} \sim 5\times10^{6} h^{-1}M_\odot\). 
Instead, clear trends emerge in both the stellar and gas content. 
Increasing $T_\mathrm{SN}$ systematically leads to higher stellar masses and lower total gas masses, whereas larger values of $t_0^{\star}$ produce the opposite behavior, resulting in lower stellar masses and larger gas reservoirs. 
Higher values of $\kappa_w$ are also associated with lower stellar masses, while the correlation with the total gas mass is weaker and mainly reflected in a reduced scatter toward larger values of the parameter.
These results indicate that the dominant feedback parameters modify the balance between stellar growth and the gas reservoir in different ways.
Since the quantity shown corresponds to the total gas mass rather than the cold star-forming component, variations in gas mass alone do not necessarily translate directly into changes in the quenched fraction. 
Instead, the quenching response likely depends on how each parameter regulates both the amount of gas available and its thermodynamic state.
Overall, despite these systematic trends, the corresponding variations in the quenched fraction remain relatively modest for low-mass galaxies, typically below $\sim10\%$ across the VTNG realizations.
The high stellar mass regime exhibits a markedly different behavior. 
Here, the dominant parameters correlate with the mean black hole mass, indicating that AGN activity becomes the primary regulator of galaxy evolution.
Increasing $\epsilon_{\rm f,high}$ substantially decreases the mean black hole mass while producing a comparatively weak increase in the mean stellar and gas mass.
At the same time, it reduces the quenched fraction, consistent with a self-regulating AGN feedback loop in which more efficient energy injection limits black hole growth while suppressing subsequent quenching \citep{Booth_2010, Bower_2017, Piotrowska_2022}. 
By contrast, increasing $T_\mathrm{SN}$ is associated with larger black hole masses and higher QF, while the mean stellar mass remains approximately unchanged. 
The absence of a clear trend with stellar mass suggests that the increase in black hole mass is not simply a consequence of more massive galaxies, but reflects a more direct response of black hole growth to changes in the feedback prescription. 
At the same time, increasing $T_\mathrm{SN}$ systematically reduces the total gas mass, indicating that changes in the gas reservoir accompany the enhanced quenching.
These results are consistent with an indirect coupling between SN feedback, gas supply, and black hole growth \citep[e.g.,][]{Dubois_2015, Torrey_2020}.
The trends associated with $\epsilon_m$ remain comparatively weak in Fig. \ref{fig:scatter} in the high stellar mass regime.
This does not contradict the RF/SHAP analysis of Sec. \ref{sec:feed_param}, where $\epsilon_m$ becomes one of the dominant parameters when the quenched fraction is analyzed as a function of black hole mass (middle panel in Fig. \ref{fig:rf_importance}). 
This difference highlights that the influence of $\epsilon_m$ is more closely linked to black hole properties than to stellar mass alone.
While the stellar mass analysis identifies $\epsilon_{\rm f,high}$ as the dominant AGN-related parameter, the black hole mass analysis reveals a stronger role for $\epsilon_m$, indicating that the relative importance of AGN feedback parameters depends on the physical variable used to characterize galaxy evolution.
Overall, unlike the low-mass regime where quenching is most closely linked to the gas reservoir, high-mass galaxies show a much stronger connection between quenching and black hole growth, reinforcing the transition from stellar and SN feedback dominated regulation to AGN-driven quenching inferred from the RF analysis.
%

\section{Centrals \& Satellites}
\label{sec:central_and_satellites}
In the previous section we showed that combining RF/SHAP analysis allows us to identify the main galaxy formation parameters driving the quenching of galaxies, and that the relative influence of each parameter changes systematically with the stellar mass range considered.
However, this mass dependence need not arise solely from the differential impact of distinct astrophysical processes on galaxies of different size and morphology. 
It may also reflect differences in their formation histories and, importantly, changes in the galaxy population with stellar mass. 
In particular, the fraction of satellite galaxies increases toward lower stellar masses, while central galaxies become increasingly dominant at higher masses \citep{Wetzel_2012,Wetzel_2013,Peng_2015}.
Motivated by this, we now analyze central and satellite galaxies separately to disentangle the effects of stellar mass and environment.
If environment, rather than stellar mass itself, were the primary driver of the trends identified above, the parameter rankings obtained in Sec. \ref{sec:robustness} could partly reflect a central or satellite dichotomy rather than a genuine mass-dependent transition in feedback physics.
To disentangle these two effects, in this section we split our galaxy sample into centrals and satellites and repeat the main analysis separately for each population, allowing us to test whether the parameter dependencies identified above persist within a fixed environmental class.

\begin{figure}[h!] 
\centering
\includegraphics[width=0.9\columnwidth]{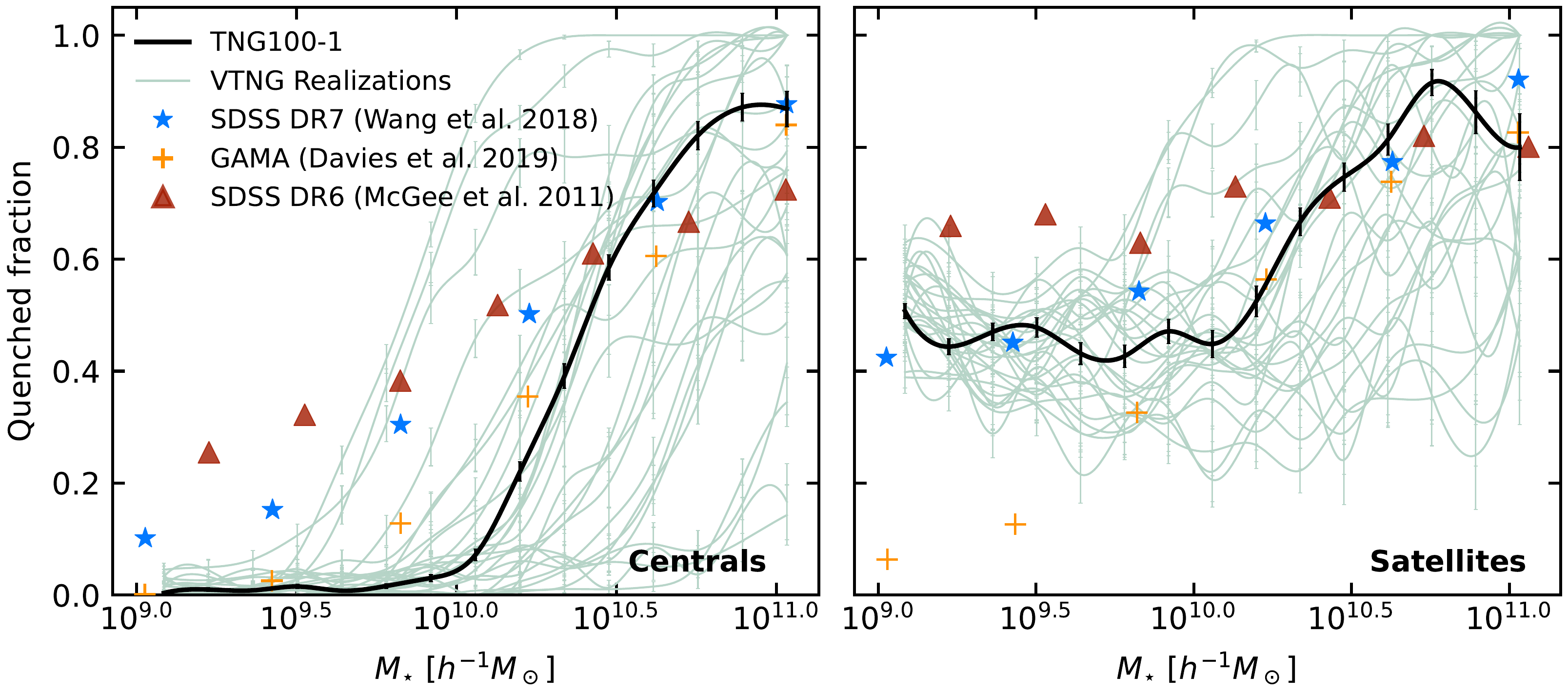}
\caption{Quenched fraction as a function of the total stellar masses of the galaxies. All the lines have been smoothed, and the error bars indicate the standard deviation across the sample. The gray lines correspond to VTNG realization and the black solid line to the reference TNG100-1 simulation. The observational measures are shown as blue stars (SDSS DR7), orange crosses (GAMA), and, red triangles (SDSS DR6 field). \textbf{Left:} Central galaxies. \textbf{Right:} Satellite galaxies.} 
\label{fig:qgf_cen_sat} 
\end{figure}

\begin{figure*}[h!] 
\centering
\includegraphics[width=0.67\textwidth]{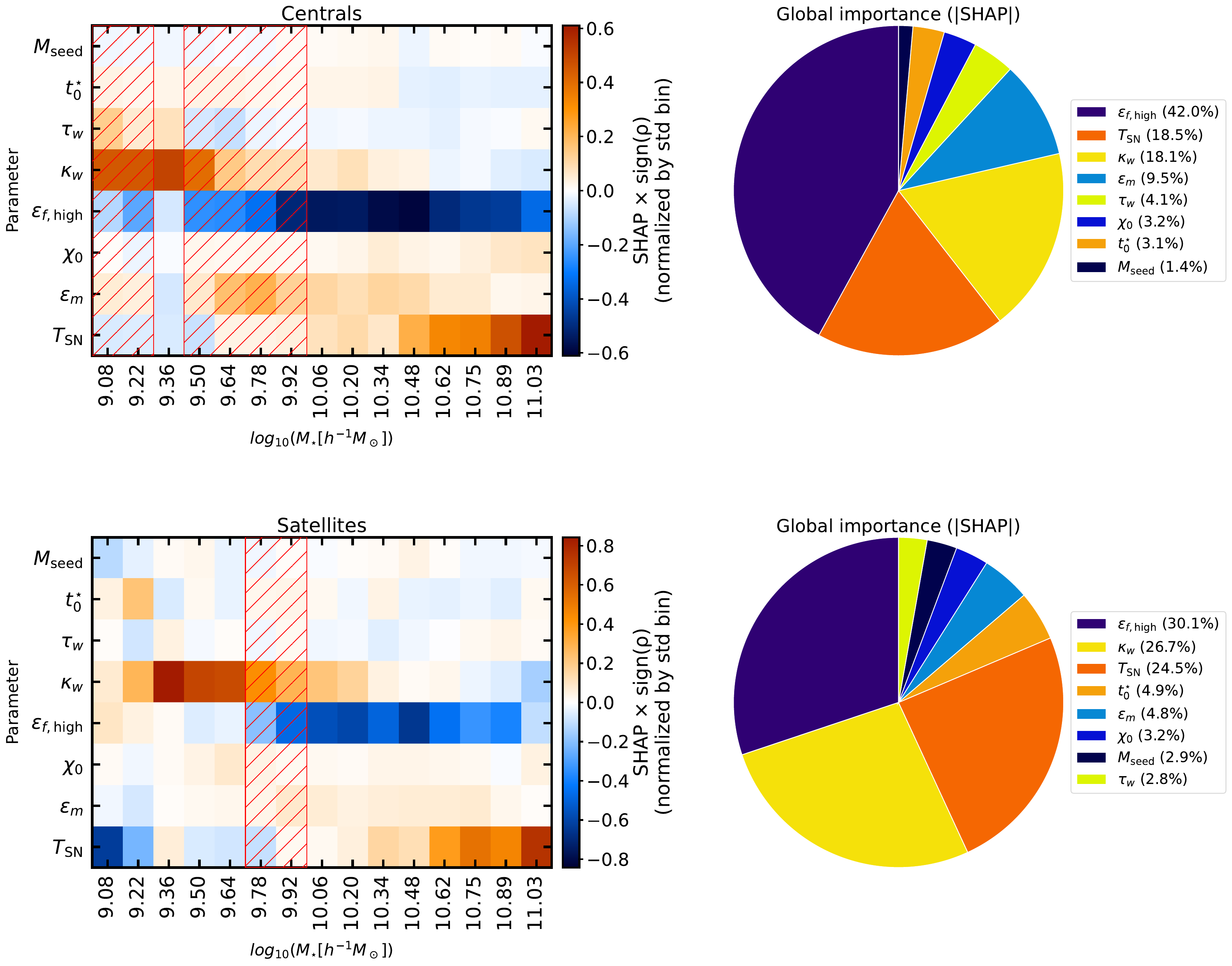}
\caption{SHAP analysis performed on the RF models of the quenched central (Top) and satellite (Bottom) galaxy fractions in the VTNG realizations. The figure follows the same format as Fig. \ref{fig:rf_importance}.} 
\label{fig:shap_cen_sat} 
\end{figure*}

The total number of central and satellite galaxies in each simulation analyzed is listed in Table~\ref{tab:galaxy_counts}.
Across the VTNG realizations, central galaxies typically account for $52.54$--$65.44\%$ of the galaxy population, while satellites represent $34.56$--$47.46\%$.
For this analysis, both populations are binned in the same stellar mass bins, and the quenched fraction is computed independently for centrals and satellites in each bin, following the definition given in Sec. \ref{sec:quenched_frac}.
The resulting QF are shown in Figure \ref{fig:qgf_cen_sat}, with central galaxies in the left panel and satellites in the right panel. 
As in Fig. \ref{fig:qgf_results}, the black curve denotes the TNG100-1 simulation, while the gray curves correspond to the VTNG realizations. 
The observational measurements from the SDSS DR7 \citep{Wang_2018}, GAMA \citep{Davies_2019}, and SDSS DR6 field \citep{McGee_2011} surveys are also included using the same symbols and definitions adopted previously.
The quenched fraction increases with stellar mass for both central and satellite galaxies, consistent with the trends obtained for the full galaxy sample.
However, the two populations exhibit distinct quenching efficiencies at fixed stellar mass, reflecting the different physical mechanisms that regulate star formation in each environment \citep{vBosch_2008, Peng_2010, Wetzel_2012}. 
Satellites generally display larger QF at low stellar masses, consistent with environmental processes such as gas stripping and starvation, whereas centrals dominate the quenched population at the high-mass end, where AGN feedback is expected to play a major role.
Both populations exhibit substantial variations among the different VTNG realizations, demonstrating that the quenching properties of centrals and satellites remain sensitive to the underlying galaxy formation parameters even after controlling for environment. 
Having computed the quenched fraction for central and satellite galaxies separately, we repeat the RF/SHAP analysis described in Sec. \ref{sec:drivers}, focusing on stellar mass. 
As before, the analyzed quantity is $\Delta \mathrm{QF}$, defined relative to the fiducial TNG100-1 simulation. 
The results are summarized in Figure \ref{fig:shap_cen_sat}.
The overall picture remains unchanged. 
In both central and satellite galaxies, quenching is governed by the same three parameters identified for the full sample: $\epsilon_{\rm f,high}$, $\kappa_w$, and $T_\mathrm{SN}$. 
Likewise, the transition from stellar-feedback dominated quenching at low stellar masses ($\log_{10}(M_\star [h^{-1}M_\odot])<9.7$) to AGN-feedback dominated quenching at higher masses is preserved in both populations. 
This indicates that the transition identified in Sec. \ref{sec:robustness} is primarily driven by galaxy mass rather than by the relative mix of central and satellite galaxies.
The main difference lies in the relative importance of the dominant parameters. 
For central galaxies, $\epsilon_{\rm f,high}$ retains a significant influence across nearly the entire stellar mass range, including the low-mass regime where $\kappa_w$ is the primary driver. 
In contrast, satellite galaxies show a weaker contribution from $\epsilon_{\rm f,high}$ and a correspondingly stronger and more extended influence of $\kappa_w$ toward higher stellar masses. 
Consistently, the global importance of $\epsilon_{\rm f,high}$ decreases by approximately 12\% from centrals to satellites, largely balanced by increased contributions from $\kappa_w$ and $T_\mathrm{SN}$. 
This shift suggests that stellar and SN feedback play a relatively larger role in regulating quenching in satellite galaxies, whereas AGN feedback remains comparatively more important for central systems.

\section{Discussion}
\label{sec:discussion}
Our previous analysis confirms the emerging picture that galaxy quenching is governed by distinct feedback mechanisms across galaxy mass.
However, by exploiting the controlled VTNG framework together with interpretable ML techniques, we move beyond simply identifying the dominant physical processes and directly determine which sub-grid parameters regulate the transition between stellar-feedback and AGN-dominated quenching. 
The RF/SHAP analysis consistently shows that only a small subset of the eight explored parameters accounts for most of the variance in the quenched fraction. 
The dominant parameters change systematically with stellar mass and, to a lesser extent, with galaxy type (central or satellite). 
Rather than acting independently, the different feedback channels dominate distinct galaxy populations, revealing how variations in the sub-grid prescriptions propagate into observable galaxy properties.
At the low stellar mass end, our results are consistent with the current theory in which stellar feedback regulates galaxy evolution by preventing efficient star formation in shallow potential wells \citep[e.g.,][]{Terrazas_2020, Zinger_2020}. 
However, the VTNG framework allows us to identify the specific sub-grid parameters responsible for this regulation. 
Among the explored parameters, the stellar feedback wind velocity parameter ($\kappa_w$) and the supernova temperature parameter ($T_{\rm SN}$) emerge as the primary drivers of the quenched fraction.
In contrast, AGN-related parameters play only a minor role in this regime. 
The additional analysis of the mean stellar, black hole, and gas masses provides further insight into these trends. 
The lack of significant correlation between the dominant parameters and black hole mass confirms that AGN activity plays only a minor role in this stellar mass range.
Instead, the stellar-feedback parameters modify both the stellar content and the total gas reservoir in different ways. 
Since the quantity analyzed here corresponds to the total gas mass rather than the cold star-forming component, these correlations should not be interpreted as a direct measure of the gas available for star formation. 
Rather, they indicate that the efficiency of stellar feedback cannot be understood solely in terms of the total gas content, but likely also depends on how feedback redistributes and thermally processes the gas.
This interpretation is broadly consistent with theoretical models in which stellar feedback drives a burst of star formation and repeated gas cycling in low-mass galaxies \citep[e.g.,][]{Oppenheimer_2010, Muratov_2015}.
The high stellar mass regime exhibits a markedly different behavior.
Previous observational and theoretical studies have consistently identified black hole mass as one of the strongest empirical predictors of galaxy quenching \citep[][]{Terrazas_2020,Davies_2020,Ayromlou_2021,Bluck_2024}, reflecting the cumulative impact of AGN feedback during galaxy evolution. 
In particular, $\epsilon_{\rm f,high}$ emerges as the dominant AGN-related parameter, producing the strongest variations in both the mean black hole mass and the quenched fraction.
This result suggests that the efficiency of high-accretion AGN feedback is the principal regulator of quenching in stellar massive galaxies within the explored parameter space. 
At the same time, increasing $\epsilon_{\rm f,high}$ reduces the quenched fraction, consistent with a self-regulating AGN feedback loop in which more efficient energy injection limits black hole growth and suppresses subsequent quenching \citep{Booth_2010, Bower_2017, Piotrowska_2022}.
Interestingly, this picture changes when quenched galaxies are analyzed as a function of black hole mass rather than stellar mass.
While $\epsilon_{\rm f,high}$ dominates the stellar mass analysis, the RF/SHAP analysis reveals that $\epsilon_m$ becomes one of the most important parameters, and its relative importance increases significantly for galaxies hosted by massive black holes.
This behavior highlights the complementary information provided by the SHAP analysis.
The result suggests that different AGN feedback parameters become relevant once galaxies are compared at similar stages of black hole growth, emphasizing that the inferred importance of the feedback prescriptions depends on the physical variable used to characterize galaxy evolution.
This interpretation is broadly consistent with the current picture in which black hole mass traces the cumulative energy released by AGN feedback and therefore provides a better proxy for quenching in massive galaxies than stellar mass alone.
Within this framework, the increasing importance of $\epsilon_m$ indicates that, once galaxies are matched at fixed black hole mass, the efficiency of the kinetic feedback mode becomes a key factor regulating whether galaxies remain quenched, whereas $\epsilon_{\rm f,high}$ primarily determines how efficiently black holes grow to reach that stage.
At the low black hole mass end, the weaker dependence on AGN-related parameters suggests that quenching may instead be more strongly influenced by processes affecting the surrounding gas reservoir. Although our analysis does not directly test these mechanisms, this scenario is qualitatively consistent with temporary quenching episodes driven by gas removal and subsequent recycling, rather than by sustained AGN activity.

Beyond the specific results for galaxy quenching, this work illustrates the potential of combining controlled simulation suites with interpretable machine-learning techniques. 
Unlike broad parameter-space explorations such as CAMELS or FLAMINGO, where cosmology, hydrodynamics, and sub-grid physics may vary simultaneously, the VTNG framework isolates the effect of individual feedback parameters while keeping the initial conditions fixed. 
Coupled with SHAP analysis, this enables a direct physical interpretation of how modifications to the feedback prescriptions propagate into observable galaxy properties, providing a complementary perspective for constraining sub-grid models. 
Previous works identified which physical process dominates galaxy quenching \citep[e.g.,][]{Ni_2023,Lim_2025}.
Our work identifies which parameters of that physical process are mainly responsible.

The comparison with observational measurements should also be interpreted with some caution. 
Although the VTNG analysis provides a physically motivated framework for relating discrepancies in the quenched fraction to variations in the underlying feedback prescriptions, the available observational data sets do not always provide a unique reference.
In particular, the SDSS and GAMA measurements exhibit systematic differences at low stellar masses, illustrating that part of the tension between simulations and observations arises from differences in galaxy classification and sample selection rather than exclusively from the physical implementation of feedback.
Similar limitations apply to the comparison of the simulated black hole and gas properties with observations.
Observational estimates of black hole masses are generally indirect and can exhibit substantial scatter depending on the adopted method and available observables \citep[e.g.,][]{Piotrowska_2022,Bluck_2023}, while the total gas mass considered here is not directly accessible observationally.
Specifically, observations such as those from ALFALFA primarily trace the atomic component of the gas, rather than the total gas reservoir considered in our analysis. 
A more direct comparison would therefore require separating the cold gas phases and considering quantities such as the cold gas to stellar mass ratio, as explored in observational studies of local galaxies and AGNs \citep[e.g.,][]{Yesuf2020a, Yesuf2020b}. 
A detailed investigation along these lines, including the connection between quenching, the different gas phases, and black hole growth, is beyond the scope of the present work but represents an interesting direction for future studies.
Within this context, the VTNG framework remains valuable because it identifies which feedback parameters are most likely to account for the observed discrepancies.
However, not all differences between the simulations and the observations should necessarily be interpreted as deficiencies of the feedback model. 
A detailed comparison between the different stellar mass observational samples, together with the best-matching VTNG realizations for each case, is presented in Appendix \ref{app:observation}.
%

\section{Conclusions}
\label{sec:conclusions}
In this work, we used the VTNG simulation suite to investigate how variations in the galaxy formation and evolution parameters affect the quenched galaxy fraction at $z=0$. 
By combining the controlled exploration of the eight VTNG feedback parameters with RF regressors and SHAP analysis, we quantified the relative importance and direction of the physical processes that regulate galaxy quenching across different stellar mass, black hole mass, and gas mass regimes.
The analysis was performed for the full galaxy population and separately for central and satellite galaxies, allowing us to test whether the mass-dependent trends persist separately for central and satellite populations.
Our main results can be summarized as follows:
\begin{itemize}
\item The quenched fraction exhibits a large variation among the VTNG realizations, demonstrating that galaxy quenching is highly sensitive to the adopted sub-grid feedback prescriptions.
\item The RF/SHAP analyses reveal that only a small subset of the eight VTNG parameters dominates the variance of the quenched fraction. The most important parameters are $\kappa_w$, $T_{\rm SN}$, $\epsilon_{\rm f,high}$, and $\epsilon_m$, while the remaining parameters have a comparatively minor impact.
\item The dominant feedback mechanisms depend strongly on stellar mass. Low-mass galaxies are primarily regulated by stellar feedback, with $\kappa_w$ emerging as the principal driver of the quenched fraction, whereas intermediate- and high-mass galaxies become increasingly sensitive to AGN-related parameters, particularly $\epsilon_{\rm f,high}$. In both mass extremes, the supernova temperature parameter $T_{\rm SN}$ has an important contribution.
\item The SHAP analysis provides not only the relative importance of the varied parameters but also the direction of their influence, allowing us to identify which changes in the sub-grid feedback prescriptions increase or decrease the quenched fraction across different stellar mass regimes.
\item Low-mass galaxies show little connection between quenching and black hole growth, whereas the high-mass regime exhibits a strong correlation with black hole mass, consistent with AGN becoming the primary regulator of galaxy evolution.
\item The relative importance of the AGN parameters depends on the physical variable used to characterize the galaxy population. While $\epsilon_{\rm f,high}$ dominates the stellar mass analysis, $\epsilon_m$ becomes significantly more important when the quenched fraction is analyzed as a function of black hole mass, highlighting the complementary information provided by different mass tracers.
\item The separation between central and satellite galaxies demonstrates that the stellar mass-dependent transition identified in the full sample remains present in both populations. Although the same dominant parameters regulate quenching in centrals and satellites, their relative importance differs, with central galaxies exhibiting a stronger sensitivity to AGN-related parameters and satellites showing a comparatively larger contribution from stellar feedback.
\item Comparisons with observations show that VTNG can link discrepancies in the quenched fraction to specific feedback prescriptions, while differences among observational data sets caution against attributing all tensions to the sub-grid model.
\end{itemize}
Overall, our results demonstrate that the combination of controlled simulation suites and interpretable ML techniques provides a powerful framework for understanding galaxy quenching. 
The methodology developed here can be naturally extended to other galaxy properties and future simulation suites, offering a promising route toward physically interpretable calibration of galaxy formation models.
%

\begin{acknowledgements}
This work has been partially supported with grants from Agencia Nacional de Investigación y Desarrollo (ANID) and the Chinese Academy of Sciences (CAS). We acknowledge the use of the High Performance Computing Resource in the Core Facility for Advanced Research Computing at the Shanghai Astronomical Observatory. We also thank Bocheng Zhu for valuable assistance with our VTNG simulations. ADMD acknowledges support from the Universidad Técnica Federico Santa María through the Proyecto Interno Regular \texttt{PI\_LIR\_25\_04}. HG is supported by the National Natural Science Foundation of China (12595313), the National SKA Program of China (2025SKA0150100), and the CAS Project for Young Scientists in Basic Research (YSBR-092). LCH was supported by the National Science Foundation of China (12233001) and the China Manned Space Program (CMS-CSST-2025-A09).
Figures were developed using \textsc{Matplotlib} \citep{Hunter_2007}. 
\end{acknowledgements}

\bibliographystyle{aa}
\bibliography{references}

\appendix
\section{Validation of the Random Forest Architecture and Simulation Sample Size}
\label{app}
\begin{figure*}[h!] 
\centering
\includegraphics[width=\textwidth]{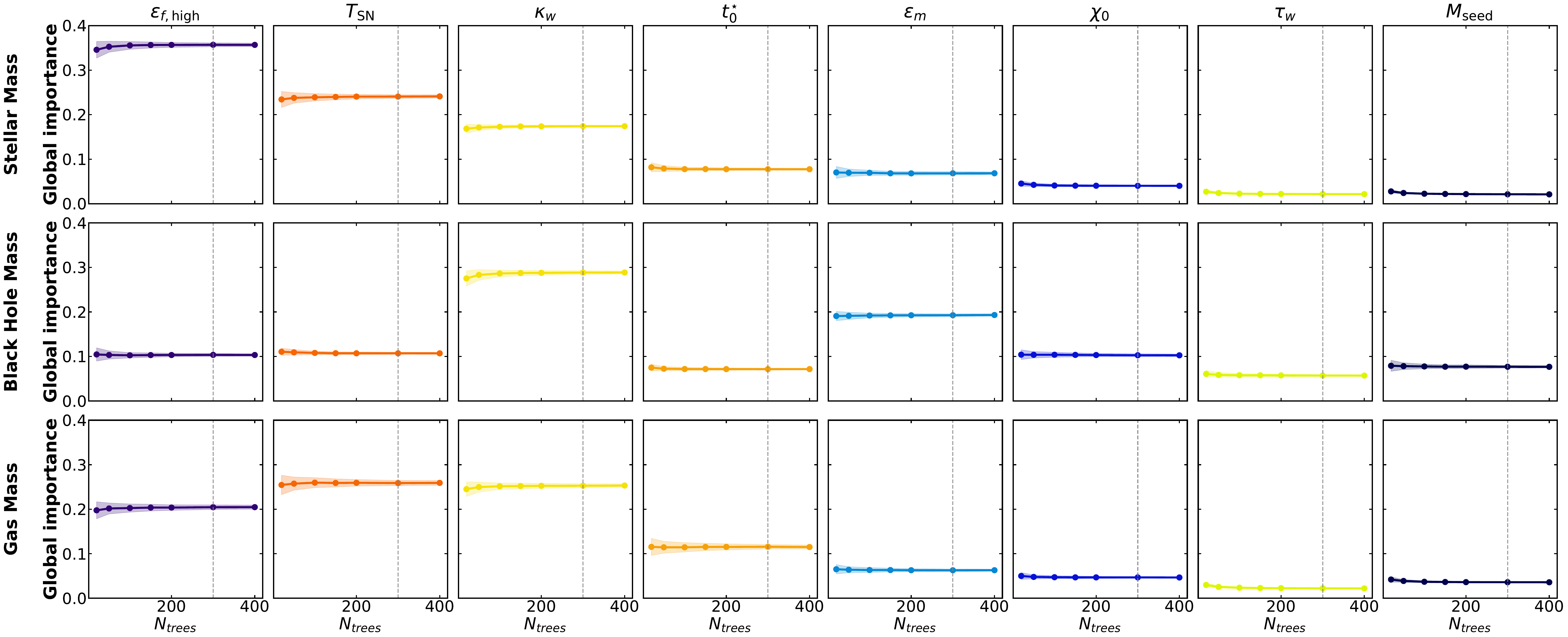}
\caption{Stability of the global SHAP importance estimates as a function of the number of decision trees $N_\mathrm{trees}$ in the Random Forest ensemble. The full dataset of 26 simulations is used throughout. For each value of $N_\mathrm{trees}$, the RF is trained 100 times with different random seeds, and the global SHAP importance of each parameter is computed for each instantiation. Solid lines show the mean importance across the 100 repetitions, and shaded bands indicate the $\pm 1\sigma$ interval, capturing the variability introduced by the stochasticity of the bootstrap sampling and random feature selection internal to the RF algorithm. Each panel corresponds to one of the VTNG parameters and each row to a different mass proxy. The importance estimates converge rapidly, and the band narrows to a negligible width for $N_\mathrm{trees} \geq 300$ (vertical dashed line), which is adopted as the default throughout this work. \textbf{Top:} Total stellar mass. \textbf{Middle:} Total black hole mass. \textbf{Bottom:} Total gas mass.} 
\label{fig:app_ntrees} 
\end{figure*}
\begin{figure*}[h!] 
\centering
\includegraphics[width=\textwidth]{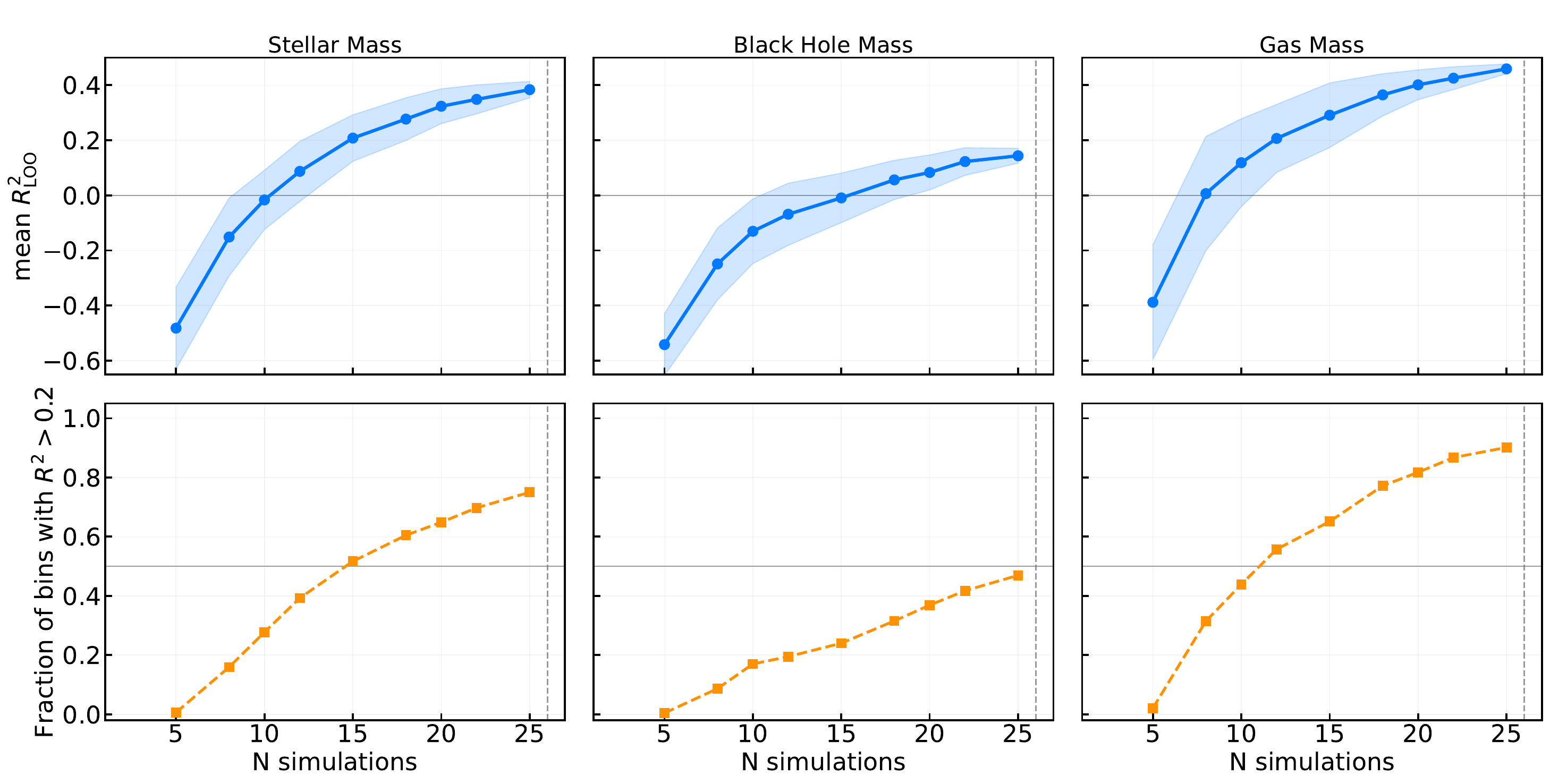}
\caption{Learning curve of RF models. The upper panels show the mean $R_{\rm LOO}^2$ as a function of $N_{\mathrm{sub}}$, while the lower panels show the corresponding fraction of mass bins with $R_{\rm LOO}^2 > 0.2$. \textbf{Left:} Total stellar mass. \textbf{Middle:} Total black hole mass. \textbf{Right:} Total gas mass.} 
\label{fig:r2_robust} 
\end{figure*}

The RF/SHAP analysis presented in this work depends on both the adopted RF architecture and the size of the available simulation suite. 
We therefore assess how the results change when varying the number of decision trees in the RF and the number of simulations used for training.
We first examine the stability of the global SHAP importance estimates as a function of the number of trees in the forest. 
We then investigate the dependence of the same importance estimates on the number of simulations, assessing whether the inferred parameter ranking is sensitive to the particular subset of simulations used. 
Finally, and most importantly, we evaluate how the predictive performance of the RF models, quantified by $R_{\rm LOO}^2$, changes with the number of simulations. 
Together, these tests assess both the stability of the inferred parameter dependencies and the predictive reliability of the models given the limited size of our simulation suite.
The RF algorithm is inherently stochastic: each tree in the ensemble is trained on a distinct bootstrap resample of the training set, and at each node split a random subset of features is evaluated.
Given the small number of available simulations, we do not perform an exhaustive hyperparameter optimization. 
Such a procedure would require tuning the model within the LOO-CV framework, leading to a nested LOO scheme in which hyperparameters are optimized on an already very limited sample. 
In this regime, the additional flexibility introduced by hyperparameter tuning could result in fitting the hyperparameters to sampling noise, potentially outweighing any gain in predictive performance. 
Moreover, for an RF with a sufficiently large number of trees, much of the estimator variance is already reduced through ensemble averaging, while parameters such as the maximum tree depth or minimum number of samples per leaf are expected to have a secondary effect relative to the uncertainty associated with the limited simulation sample.
We therefore focus on the number of trees as the main architectural choice and assess the stability of the SHAP importance estimates with respect to this parameter. 
For a fixed dataset, the variability of the importance estimates across independent RF instantiations reflects the residual stochasticity of the estimator, which is expected to decrease as more trees are added to the ensemble.
To quantify this effect, we train 100 independent RF models on the complete set of 26 simulations for each number of estimators $N_{\mathrm{trees}} \in {20, 50, 100, 150, 200, 300, 400}$, each time using a different random seed while keeping the training data fixed. 
For each instantiation, we compute the global SHAP importance of every parameter following the procedure described in Sec. \ref{sec:feed_param}. 
The mean and standard deviation of the resulting importance distributions are shown in Figure \ref{fig:app_ntrees} as a function of $N_{\mathrm{trees}}$, separately for each of the VTNG parameters.
The results show that for small ensembles ($N_\mathrm{trees} \lesssim 50$) the importance estimates exhibit a non-depreciable scatter across random seeds, reflecting the high sensitivity of individual trees to their particular bootstrap sample.
As $N_\mathrm{trees}$ increases, the standard deviation narrows rapidly. 
For $N_\mathrm{trees} \geq 300$, the band of mean $\pm 1\sigma$ is narrow for all parameters, and additional trees produce no meaningful change in either the rankings or the relative importance. 
We therefore adopt $N_\mathrm{trees} = 300$ as the default throughout this work, a value that lies comfortably within the convergence plateau while remaining computationally efficient.
\begin{figure*}[h!] 
\centering
\includegraphics[width=\textwidth]{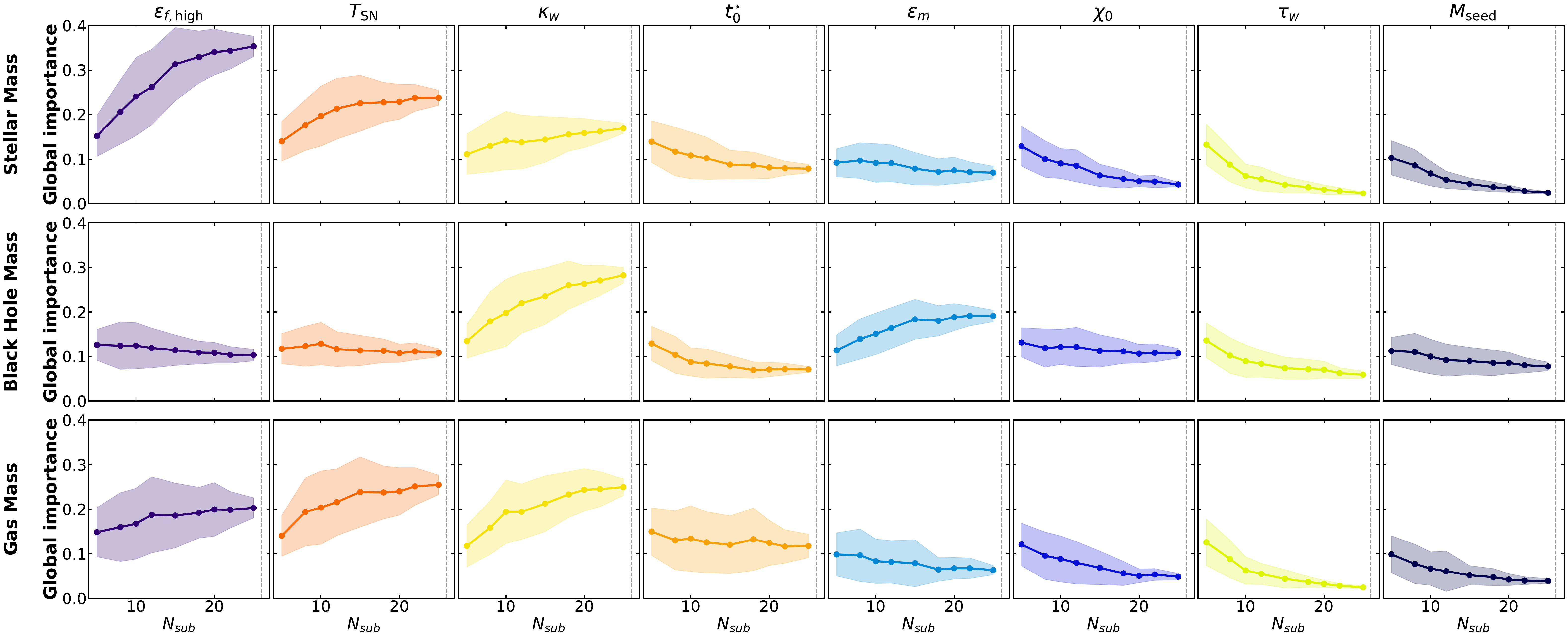}
\caption{Robustness of the global SHAP importance estimates to the number of simulations $N_\mathrm{sub}$ used to train the Random Forest. The number of trees is fixed at $N_\mathrm{trees}=300$ throughout. For each value of $N_\mathrm{sub}$, a total of $K=100$ random subsets of size $N_\mathrm{sub}$ are drawn without replacement from the full suite of 26 simulations; the RF/SHAP pipeline is applied to each subset independently, and the global SHAP importance is recorded for each parameter. Solid lines show the mean importance across the 100 repetitions, while shaded bands indicate the $\pm1\sigma$ interval, quantifying the sensitivity of the importance estimate to the particular subset of simulations selected from the available suite. Each panel corresponds to one of the VTNG parameters for each of the three mass proxies (rows). The vertical dashed line marks the full suite size ($N_\mathrm{sub}=26$). \textbf{Top:} Total stellar mass. \textbf{Middle:} Total black hole mass. \textbf{Bottom:} Total gas mass.} 
\label{fig:app_nsims} 
\end{figure*}
A complementary question is whether the inferred parameter importance is robust to the particular composition of the available simulation suite. 
In particular, a small number of simulations with atypical parameter combinations or responses could disproportionately influence the global SHAP importance. 
We therefore test the internal stability of the inferred importance rankings by repeatedly applying the RF/SHAP analysis to random subsets of the 26 available simulations.
This test should not be interpreted as an assessment of how the predictive performance or parameter ranking would improve with a larger simulation suite. 
Rather, it asks whether the conclusions drawn from the current sample are dominated by a small number of individual simulations. 
Since the subsets are drawn from the same pool of 26 simulations, this constitutes an internal convergence test: as $N_{\mathrm{sub}}$ approaches 26, the results are expected by construction to converge towards those obtained from the full sample. 
The test therefore provides a measure of robustness to the particular subset of the available simulations, but does not establish how the inferred importance ranking would change for a substantially larger suite of simulations.
To perform this test, we repeat the full RF/SHAP pipeline for each of $K=100$ random subsets of size $N_{\mathrm{sub}}$, drawn without replacement from the 26 available simulations, for each $N_{\mathrm{sub}} \in {5, 8, 10, 12, 15, 18, 20, 22, 25}$. 
For each subset and parameter, we compute the global SHAP importance and record the mean and standard deviation of the resulting distributions across the 100 repetitions.
The results are shown in Figure~\ref{fig:app_nsims}, with one panel per parameter for each of the three mass proxies.
The band of mean $\pm 1\sigma$ is broad at small $N_\mathrm{sub}$, indicating that the estimated importance is sensitive to the particular subset of simulations when only a few realizations are considered.
As $N_\mathrm{sub}$ increases, the mean converges toward the value obtained with the full dataset, marked by the vertical dashed line at $N_\mathrm{sub}=26$, while the standard deviation decreases, indicating increasing stability against the particular subset of realizations selected from the available simulation suite.
These results show that, within the current set of 26 simulations, the inferred parameter importance becomes increasingly stable as more realizations are included and is not strongly dominated by a small subset of simulations.
To last, we investigate how the predictive performance of the RF models depends on the number of simulations available for training. 
This provides a complementary assessment of the strengths and limitations of the current simulation suite.
For each of the three mass proxies, we repeat the LOO-CV analysis for random subsets of $N_\mathrm{sub} \in {5, 8, 10, 12, 15, 18, 20, 22, 25}$ simulations, using 100 independent subsets for each sample size. 
For each subset, the $R_{\rm LOO}^2$ values are computed independently for all stellar mass bins and then summarized in two complementary ways. 
First, we calculate the mean $R_{\rm LOO}^2$ across mass bins, providing an overall measure of the predictive performance at a given sample size. 
Second, we calculate the fraction of mass bins with $R_{\rm LOO}^2 > 0.2$, quantifying the fraction of the parameter space for which the RF provides a predictive model with a comparatively reliable performance. 
The mean and dispersion of these quantities across the 100 random subsets are used to characterize their dependence on $N_\mathrm{sub}$.
The results are shown in Figure~\ref{fig:r2_robust}. 
The upper panels show the mean $R_{\rm LOO}^2$ as a function of the number of simulations, while the lower panels show the corresponding fraction of mass bins with $R_{\rm LOO}^2 > 0.2$. 
From left to right, the columns correspond to the total stellar, black hole, and gas masses, respectively.
We can observe that the predictive performance improves systematically as the number of simulations increases for all three mass proxies, although the rate of improvement decreases noticeably beyond approximately $N_\mathrm{sub}\sim20$. 
A similar trend is seen in the fraction of mass bins with $R_{\rm LOO}^2>0.2$, although its dependence on the sample size is weaker. 
These results indicate that the RF models achieve a reasonable predictive performance with the current simulation suite, while also showing that additional simulations could still provide improvements. 
In particular, the flattening of the mean $R_{\rm LOO}^2$ curves around $N_\mathrm{sub}\sim20$ suggests that the main gain in predictive performance is already achieved with roughly this number of simulations.
The total black hole mass shows the weakest predictive performance among the three mass proxies.
In this case, fewer than half of the black hole mass bins reach $R_{\rm LOO}^2>0.2$, indicating that the corresponding RF models are less reliable across the full mass range. 
One possible explanation is the smaller variation of $\Delta \rm{QF}$ with the black hole mass across the VTNG realizations, which provides less information for the model to learn from the available simulations.
In such a regime, increasing the number of simulations may be particularly beneficial for resolving the underlying dependence.
Taken together, these tests support the adopted RF configuration and its use for the goals of this work. 
The results indicate that the available simulation suite provides an adequate basis for using RF/SHAP as an exploratory tool to identify parameter dependencies and guide the subsequent physical analysis. 
However, this should not be interpreted as evidence that the current suite is sufficient for high-precision prediction. 
Applications such as the construction of accurate emulators or detailed interpolation of the parameter space would require a substantially larger and more densely sampled simulation suite. 
Our use of ML is instead intended to identify and quantify trends that can be further investigated through their physical interpretation.
%

\section{Random Forest implementation and SHAP methodology}
\label{app:rf_method}
\begin{figure}[h!] 
\centering
\includegraphics[width=1.05\columnwidth]{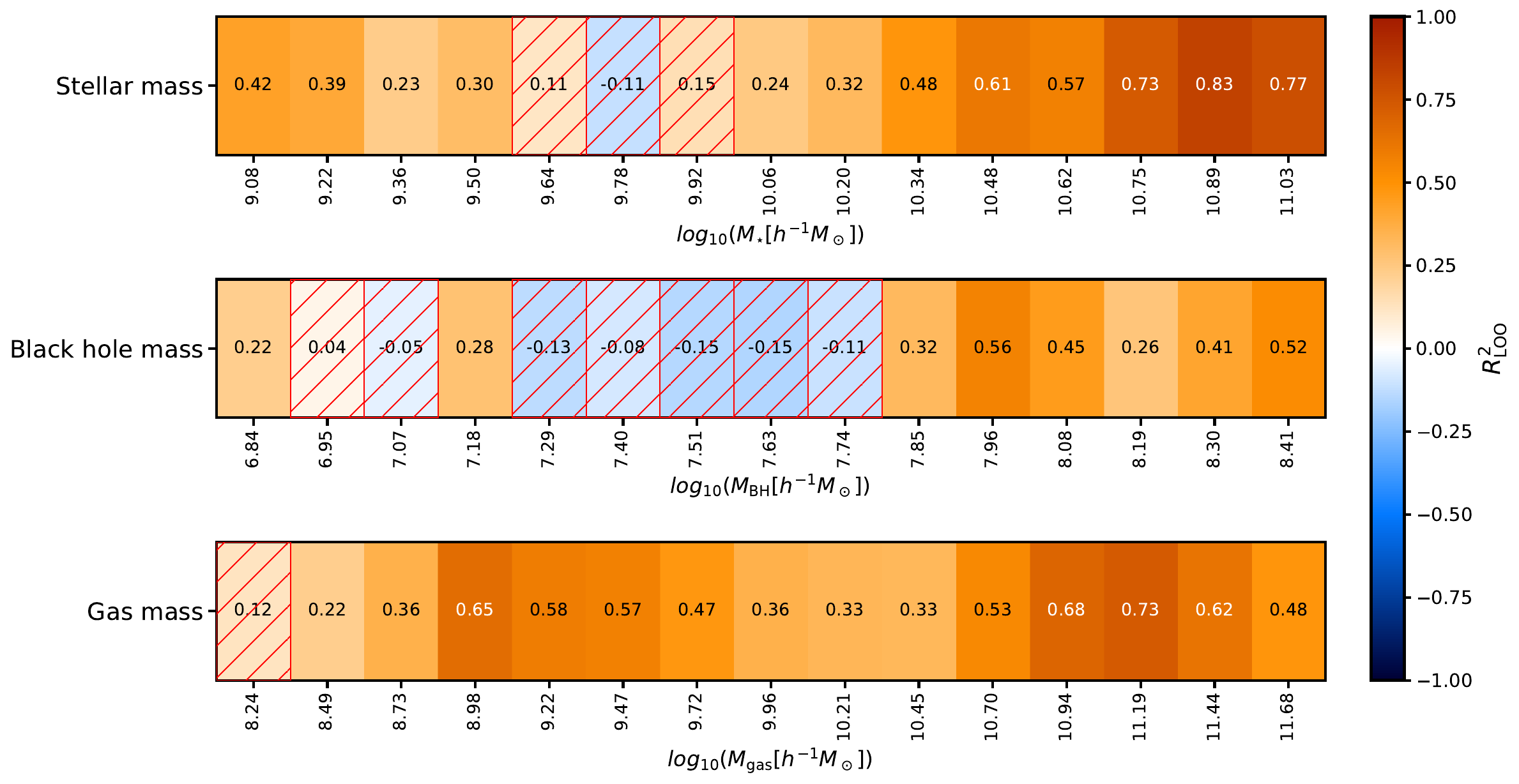}
\caption{$R_{\rm LOO}^2$ values for each mass bin resulting from the independent RF model evaluated using the LOO-CV technique. \textbf{Top:} Stellar mass bins. \textbf{Middle:} Black hole mass bins. \textbf{Bottom:} Gas mass bins.} 
\label{fig:R2_LOO_panels} 
\end{figure}

\begin{figure*}[h!] 
\centering
\includegraphics[width=0.85\textwidth]{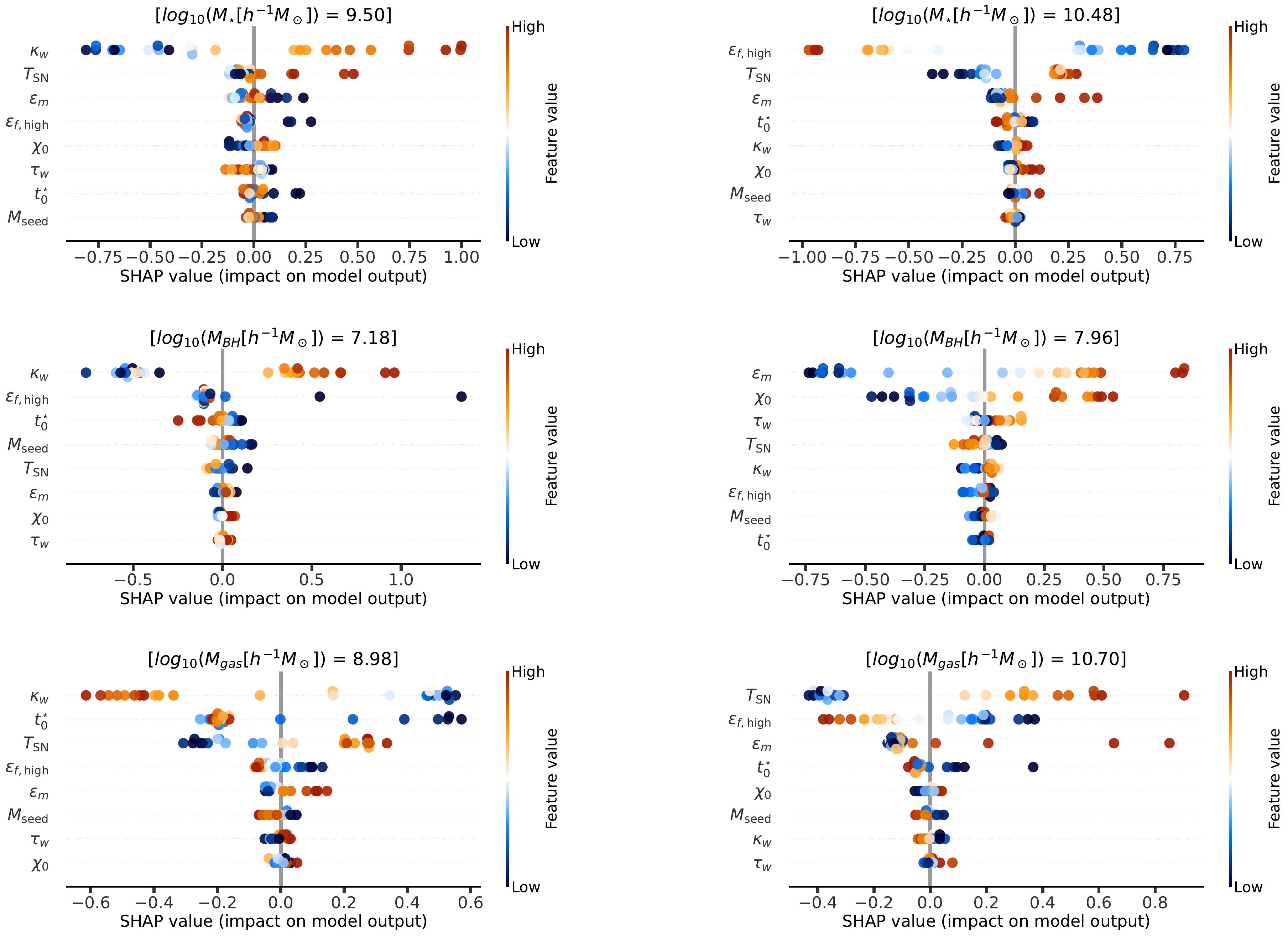}
\caption{Beeswarm representation of the SHAP values computed for different mass bins. In each panel, the points correspond to a single VTNG simulation; the color of the points encodes the relative change of the corresponding parameter in the VTNG realization with respect to the TNG100-1 value. The horizontal position of each point is given by its SHAP value. \textbf{Left:} Single low mass bin. \textbf{Right:} Single high mass bin. \textbf{Top:} Stellar mass bins. \textbf{Middle:} Black hole mass bins. \textbf{Bottom:} Gas mass bins.} 
\label{fig:shap_interpretation} 
\end{figure*}

This appendix summarizes the implementation details of the RF analysis and the SHAP-based interpretation of the model predictions. 
These technical aspects are separated from the main text to emphasize the physical interpretation of the results while providing the information required to reproduce the analysis.
As is described in Sec. \ref{sec:feed_param}, for each mass bin $k$ an independent RF model was first evaluated using the LOO-CV technique (see Sec. \ref{sec:ML_tech}), yielding the coefficient of determination $R_{\rm{LOO}}^2$ (Eq. \ref{eq:R}), which serves as a diagnostic of the model predictive performance in that mass regime.
The resulting $R_{\rm LOO}^2$ values for each mass bin are presented in Figure \ref{fig:R2_LOO_panels}, which provides a direct view of the predictive performance across the three mass proxies. 
The figure is divided into three panels, showing the stellar, black hole, and total gas mass bins from top to bottom, respectively. 
The color of each cell represents the corresponding $R_{\rm LOO}^2$ value, as indicated by the color bar, while bins with $R_{\rm LOO}^2 < 0.2$ are marked in red to highlight those for which the predictive performance is comparatively low.
Each RF model was then fitted using the full set of simulations, and SHAP values were computed with the TreeExplainer algorithm for all simulations.
This yields a matrix $\boldsymbol{\Phi}_k \in \mathbb{R}^{N_\mathrm{sim} \times N_\mathrm{params}}$, whose element $\Phi_{s,j,k}$ represents the SHAP value of parameter $j$ for simulation $s$, quantifying its contribution to the prediction of $\Delta \mathrm{QF}_{k}^{(s)}$.
The local importance of parameter $j$ in mass bin $k$ is defined as the mean absolute SHAP value over all simulations,
\begin{equation}
    |\phi_{j,k}| = \frac{1}{N}\sum_s |\Phi_{s, j, k}|
\end{equation}
The directional sign is then recovered from the Spearman rank correlation between the parameter deviations $\Delta_j^{(s)}$ and their SHAP values $\Phi_{s,j,k}$.
Figure \ref{fig:shap_interpretation} illustrates this procedure for two representative mass bins, one in the low-mass and one in the high-mass regime, for each of the three mass proxies considered. 
The figure provides a direct visualization of how the SHAP values of an individual mass bin are obtained; the same procedure is independently applied to every mass bin, yielding the parameter importances that are subsequently summarized in the heatmaps.
Each point corresponds to a single simulation and is positioned according to the SHAP value of a given parameter, which quantifies the magnitude and direction of its contribution to the predicted $\Delta \mathrm{QF}$ in the corresponding mass bin.
Positive (negative) SHAP values indicate that the parameter variation increases (decreases) the quenched fraction relative to the reference model.
Point colors encode the logarithmic parameter deviation from the TNG100-1 reference. 
To avoid overplotting, points are vertically spread according to their local density.
Parameters are ranked by decreasing mean absolute SHAP value. 
Consequently, the horizontal extent of the point distribution reflects the overall importance of each parameter, while the correlation between color and SHAP value indicates whether increasing the parameter tends to increase or decrease the quenched fraction.
The resulting signed importance matrix $\boldsymbol{\phi}^{\pm} \in \mathbb{R}^{N_\mathrm{params} \times N_\mathrm{bins}}$ (see Eq. \ref{eq:sign}) is the primary output of the analysis and is presented as heatmaps in Fig. \ref{fig:rf_importance}, where positive (negative) values indicate that increasing the parameter tends to increase (decrease) the quenched fraction in the corresponding mass bin.
To obtain a single scalar summary of each parameter's overall influence across the full mass range, the per-bin absolute SHAP values are aggregated via an unweighted mean over bins,
\begin{equation}
    \label{eq:global}
    \Phi^{global}_j = \frac{1}{N_{bins}}\sum_k|\phi_{j,k}|
\end{equation}
and subsequently normalized so that the values sum to unity, yielding a fractional global importance that can be compared directly across parameters and across mass proxies.
This global ranking is used to order parameters in the summary pie charts presented alongside each heatmap in Fig. \ref{fig:rf_importance}, which display the top contributing parameters together with their relative fractional importance.
The impact of the number of VTNG simulations used in this analysis on the results is studied in depth in App. \ref{app}.
\begin{figure}[h!] 
\centering
\includegraphics[width=1.05\columnwidth]{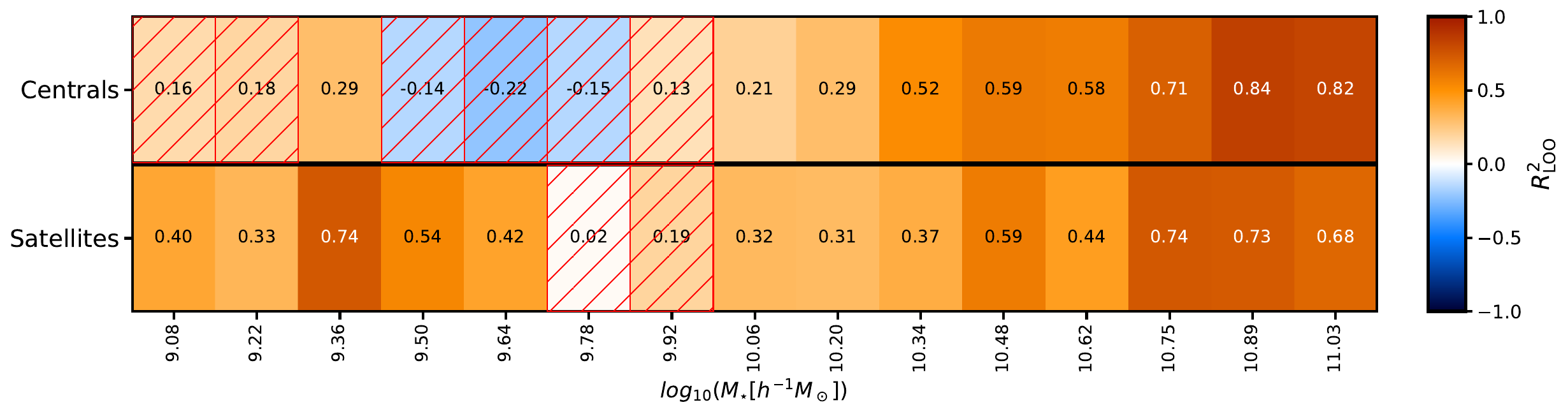}
\caption{$R_{\rm LOO}^2$ values for each stellar mass bin resulting from the independent RF model evaluated using the LOO-CV technique. \textbf{Top:} Central galaxies. \textbf{Bottom:} Satellite galaxies.} 
\label{fig:r2_loo_populations} 
\end{figure}
For completeness, we also show in Figure \ref{fig:r2_loo_populations} the $R_{\rm LOO}^2$ values obtained when the analysis is performed separately for central and satellite galaxies in Sec. \ref{sec:central_and_satellites}. 
This additional test is restricted to the stellar mass bins and follows the same procedure described above. 
The figure follows the same configuration as Fig. \ref{fig:R2_LOO_panels}, with the top panel corresponding to central galaxies and the bottom one to satellites.
This provides a complementary view of the predictive performance of the RF models across the two galaxy populations.

\section{Detailed comparison with observational data sets}
\label{app:observation}

As shown in Sects. \ref{sec:quenched_frac} and \ref{sec:central_and_satellites}, the fiducial TNG100-1 model does not reproduce all observational data sets equally well. 
Moreover, part of the disagreement arises from differences among the observational samples themselves (Figs. \ref{fig:qgf_results} and \ref{fig:qgf_cen_sat}). 
This is particularly evident at low stellar masses, where the SDSS and GAMA surveys predict different QF despite probing the nearby Universe. 
In particular, the quenched-galaxy classification adopted by \citet{Davies_2019} yields systematically lower QF below $M_\star \leq 10^{9.5}\ h^{-1}M_\odot$ than the definitions adopted in this work and by \citet{Wang_2018} and \citet{McGee_2011} for SDSS DR7 and DR6 field, respectively.  
Therefore, part of the apparent tension between simulations and observations may originate from differences in the observational classifications rather than from the physical model alone.
In this appendix, we use the VTNG suite to place these discrepancies in the context of galaxy formation physics. 
Because all VTNG realizations share the same initial conditions and differ only in the values of the subgrid galaxy-formation parameters, they provide a controlled framework to identify which feedback prescriptions drive the variations in the quenched fraction. 
Moreover, the SHAP analysis determines not only which parameters dominate the variations, but also whether increasing or decreasing each parameter raises or lowers the predicted quenched fraction. This information allows the discrepancies between TNG100-1 and the observational samples to be interpreted in terms of the underlying feedback model.
At low stellar masses, where the largest discrepancies between the SDSS and GAMA measurements are found, the quenched fraction is primarily controlled by stellar-feedback parameters, particularly $\kappa_w$.
Our SHAP analysis indicates that lower values of $\kappa_w$, corresponding to slower galactic winds, decrease the quenched fraction of low-mass galaxies, bringing the VTNG predictions closer to the GAMA measurements.
At intermediate stellar masses, the quenched fraction becomes increasingly sensitive to the AGN feedback parameter $\epsilon_{\rm f,high}$. 
In this regime, the SDSS measurements (Fig. \ref{fig:qgf_results}) generally exhibit larger QF than the fiducial TNG100-1 model. 
According to our SHAP analysis, reducing $\epsilon_{\rm f,high}$ would increase the predicted quenched fraction, leading to better agreement with these observations.
At the highest stellar masses, the trend reverses, with TNG100-1 predicting slightly larger QF than observed. 
As shown in Fig. \ref{fig:rf_importance}, this regime is mainly influenced by the SN feedback parameter $T_{\rm SN}$, suggesting that the fiducial model may overestimate either the amount of energy injected per SN event.
The same interpretation can be extended to central and satellite galaxies separately. 
Because central galaxies are expected to be less affected by environmental processes than satellites, they provide a cleaner probe of the internal feedback prescriptions implemented in the simulations.
At the same time, comparing both populations helps assess whether the remaining discrepancies originate primarily from the feedback model or from the treatment of environmental effects.
At low stellar masses, the observed quenched fraction of central galaxies, particularly in the SDSS samples, is systematically higher than that predicted by TNG100-1 (Fig. \ref{fig:qgf_cen_sat}). 
Our results suggest that larger $\kappa_w$ would increase the quenched fraction of low-mass central galaxies and improve the agreement with the observations. 
In contrast, the satellite population already reproduces the SDSS DR7 measurements of \citet{Wang_2018}, although it remains below the SDSS DR6 field sample of \citet{McGee_2011}. 
None of the VTNG realizations reaches substantially lower QF inferred from the GAMA classification, indicating that this discrepancy cannot be explained solely by the variations of the feedback parameters explored by VTNG.
At intermediate stellar masses, TNG100-1 continues to underestimate the quenched fraction of central galaxies, particularly relative to the SDSS samples. 
This suggests that a recalibration of the black hole feedback efficiency could improve the agreement with the observed abundance of passive central galaxies. 
The discrepancy is smaller for satellites, implying that the remaining differences are more likely associated with the internal feedback prescriptions than with environmental effects.
At the highest stellar masses, the agreement between TNG100-1 and the observational measurements becomes substantially better for both central and satellite galaxies, indicating that the remaining discrepancies are comparatively small in this regime.
Table~\ref{tab:obs_comparation} complements the previous discussion by listing the VTNG realization that minimizes the RMSE with respect to each observational sample. 
Despite the differences among the surveys, the preferred realizations display consistent trends. 
Those that best reproduce the GAMA measurements systematically favor lower values of $\kappa_w$, reflecting the smaller QF of low-mass galaxies inferred from this survey. 
At intermediate stellar masses, the preferred realizations generally exhibit negative deviations in $\epsilon_{\rm f,high}$, particularly for the SDSS DR6 field sample.
Finally, at the high mass end they consistently favor lower values of $T_{\rm SN}$. 
These trends reinforce the interpretation derived from the SHAP analysis.
\begin{table*}[h!] 
\centering 
\caption{Most similar VTNG realization to observational data. The parameter set of the simulation are listed as logarithmic offsets relative to TNG100-1.} 
\label{tab:obs_comparation} 
\scriptsize 
\setlength{\tabcolsep}{4pt} 
\begin{tabular}{lcccc c} 
\hline 
Realization & Observational & RMSE & Galaxy population & Parameter set \\ 
            & survey        &      &                   & $[\log_{10}(\mathrm{Par}_{\mathrm{VTNG\ box}}/\mathrm{Par}_{\mathrm{TNG}})]$ \\ 
\hline 
VTNG 01 & \shortstack[c] {GAMA \\ \cite{Davies_2019} }& 0.075 & All & 
\shortstack[l]{[$\epsilon_{f,\text{high}}=-0.9$; $T_{\text{SN}}=-0.59$; $\tau_w=-0.45$; $\chi_0=0.24$; \\ \hfill $t^{\star}_0=-0.21$; $M_{\text{seed}}=0.16$; $\epsilon_m=-0.15$; $\kappa_w=-0.06$]} \\ 
\hline
VTNG 22 & \shortstack[c] {SDSS DR7 \\ \cite{Wang_2018} } & 0.050 & All & 
\shortstack[l]{[$\tau_w=-0.89$; $T_{\text{SN}}=-0.63$; $M_{\text{seed}}=0.58$; $\epsilon_{f,\text{high}}=-0.23$; \\ $\epsilon_m=0.19$; $t^{\star}_0=-0.18$; $\kappa_w=0.11$; $\chi_0=0.01$]} \\ 
\hline
VTNG 22 & \shortstack[c] {SDSS DR6 field \\ \cite{McGee_2011} } & 0.111 & All & 
\shortstack[l]{[$\tau_w=-0.89$; $T_{\text{SN}}=-0.63$; $M_{\text{seed}}=0.58$; $\epsilon_{f,\text{high}}=-0.23$; \\ $\epsilon_m=0.19$; $t^{\star}_0=-0.18$; $\kappa_w=0.11$; $\chi_0=0.01$]} \\ 
\hline 
VTNG 01 & \shortstack[c] {GAMA \\ \cite{Davies_2019} } & 0.079 & Centrals & 
\shortstack[l]{[$\epsilon_{f,\text{high}}=-0.9$; $T_{\text{SN}}=-0.59$; $\tau_w=-0.45$; $\chi_0=0.24$; \\ \hfill $t^{\star}_0=-0.21$; $M_{\text{seed}}=0.16$; $\epsilon_m=-0.15$; $\kappa_w=-0.06$]} \\ 
\hline 
VTNG 22 & \shortstack[c] {SDSS DR7 \\ \cite{Wang_2018} } & 0.124 & Centrals & 
\shortstack[l]{[$\tau_w=-0.89$; $T_{\text{SN}}=-0.63$; $M_{\text{seed}}=0.58$; $\epsilon_{f,\text{high}}=-0.23$; \\ $\epsilon_m=0.19$; $t^{\star}_0=-0.18$; $\kappa_w=0.11$; $\chi_0=0.01$]} \\ 
\hline
VTNG 13 & \shortstack[c] {SDSS DR6 field \\ \cite{McGee_2011} } & 0.162 & Centrals & 
\shortstack[l]{[$\epsilon_{f,\text{high}}=-0.99$; $\epsilon_m=-0.67$; $\tau_w=-0.33$; $T_{\text{SN}}=-0.17$;\\ $M_{\text{seed}}=-0.09$; $\kappa_w=0.08$; $\chi_0=-0.06$; $t^{\star}_0=0.00$]} \\ 
\hline 
VTNG 02 & \shortstack[c] {GAMA \\ \cite{Davies_2019} } & 0.145 & Satellites & 
\shortstack[l]{[$\epsilon_{f,\text{high}}=-0.44$; $T_{\text{SN}}=-0.33$; $\kappa_w=-0.28$; $\tau_w=0.18$; \\ $M_{\text{seed}}=-0.15$; $t^{\star}_0=0.12$; $\epsilon_m=0.11$; $\chi_0=0.03$]} \\ 
\hline 
VTNG 17 & \shortstack[c] {SDSS DR7 \\ \cite{Wang_2018} } & 0.050 & Satellites & 
\shortstack[l]{[$\epsilon_m=-0.85$; $M_{\text{seed}}=0.72$; $\tau_w=-0.50$; $\epsilon_{f,\text{high}}=-0.25$; \\ $\chi_0=0.22$; $\kappa_w=0.06$; $t^{\star}_0=-0.04$; $T_{\text{SN}}=-0.02$]} \\ 
\hline 
VTNG 13 & \shortstack[c] {SDSS DR6 field \\ \cite{McGee_2011} } & 0.075 & Satellites & 
\shortstack[l]{[$\epsilon_{f,\text{high}}=-0.99$; $\epsilon_m=-0.67$; $\tau_w=-0.33$; $T_{\text{SN}}=-0.17$;\\ $M_{\text{seed}}=-0.09$; $\kappa_w=0.08$; $\chi_0=-0.06$; $t^{\star}_0=0.00$]} \\  
\hline 
\end{tabular} 
\end{table*}
%

\end{document}